\documentclass[lettersize,journal]{IEEEtran}

\usepackage{amsmath,amsfonts}
\usepackage{array}
\usepackage{textcomp}
\usepackage{url}
\usepackage{verbatim}
\def\BibTeX{{\rm B\kern-.05em{\sc i\kern-.025em b}\kern-.08em
    T\kern-.1667em\lower.7ex\hbox{E}\kern-.125emX}}
\usepackage{balance}

\ifCLASSOPTIONcompsoc
  \usepackage[nocompress]{cite}
\else
  \usepackage{cite}
\fi
\ifCLASSINFOpdf
  \usepackage{graphicx}
\else
\fi
\usepackage{amsmath}
\usepackage[nohyperlinks,nolist]{acronym}
\usepackage{dblfloatfix}
\usepackage{comment}

\usepackage{pgfplots}
\pgfplotsset{compat=1.18}
\usepackage{pgfplotstable}

\usepackage{rotating}
\usepackage[caption=false, labelfont={scriptsize}, textfont={scriptsize}]{subfig}

\usepackage{booktabs}
\usepackage{longtable}
\usepackage{multirow}
\usepackage{tablefootnote}

\usepackage[dvipsnames,table]{xcolor}
\usepackage{colortbl}
\definecolor{lightGray}{gray}{0.9}

\usepackage[colorlinks=true, linkcolor=blue, citecolor=Green, urlcolor=blue]{hyperref}
\usepackage{cleveref}

\usepackage{xparse}
\usepackage{xfp}

\usepackage{lipsum}
\usepackage{amssymb}

\usepackage{algorithm}
\usepackage{algpseudocode}
\algrenewcommand{\algorithmiccomment}[1]{// \textit{#1}}
\usepackage{longdivision}

\newcommand{\myCommentMAN}[1]{\textbf{\textcolor{Green}{Mario:~#1}}}

\newcommand{\positenv}[2]{Posit$\langle #1, #2 \rangle$}

\newcommand{\ra}[1]{\renewcommand{\arraystretch}{#1}}

\newcommand{\etal}{\textit{et al.}}

\ExplSyntaxOn
\NewDocumentCommand{\roundtwo}{m}
 {
  \tl_if_blank:nTF {#1}
   {} 
   {\fpeval{round(#1,2)}}
 }
\ExplSyntaxOff

\NewDocumentCommand{\area}{m O{true}}{%
    \IfEqCase{#2}{%
        {true}{#1\,\emph{$\mu m^{2}$}}%
        {false}{#1}%
    }[\PackageWarning{area}{Invalid boolean value}]%
}

\NewDocumentCommand{\power}{m O{true}}{%
    \IfEqCase{#2}{%
        {true}{#1\,\emph{$mW$}}%
        {false}{#1}%
    }[\PackageWarning{power}{Invalid boolean value}]%
}

\NewDocumentCommand{\energy}{m O{true}}{%
    \IfEqCase{#2}{%
        {true}{#1\,\emph{$pJ$}}%
        {false}{#1}%
    }[\PackageWarning{energy}{Invalid boolean value}]%
}

\NewDocumentCommand{\delay}{m O{true}}{%
    \IfEqCase{#2}{%
        {true}{#1\,\emph{$ns$}}%
        {false}{#1}%
    }[\PackageWarning{delay}{Invalid boolean value}]%
}

\NewDocumentCommand{\frequency}{m O{true}}{%
    \IfEqCase{#2}{%
        {true}{#1\,\emph{$Ghz$}}%
        {false}{#1}%
    }[\PackageWarning{delay}{Invalid boolean value}]%
}

\begin{document}

\title{High-frequency Multispeculative Multiply-Accumulation Unit for Fused Posit Arithmetic}

\author{Mario~Alonso,
        Miguel Ángel Sacristán,
        Guillermo~Botella,~\IEEEmembership{Senior,~IEEE,}
        and~Alberto~A.~Del~Barrio,~\IEEEmembership{Senior,~IEEE}%
        
\IEEEcompsocitemizethanks{\IEEEcompsocthanksitem All authors are with the Facultad de Informática, Universidad Complutense de Madrid, 28040 Madrid, Spain.\protect\\
E-mail: \{marioa25, msacri02, gbotella, abarriog\}@ucm.es}%

\thanks{Manuscript received April 19, 2005; revised August 26, 2015.}}

\markboth{Journal of \LaTeX\ Class Files,~Vol.~14, No.~8, August~2015}%
{Shell \MakeLowercase{\textit{et al.}}: Bare Advanced Demo of IEEEtran.cls for IEEE Computer Society Journals}

\IEEEtitleabstractindextext{%
\begin{abstract}

Posit arithmetic offers a compelling alternative to the IEEE 754 floating-point standard, providing enhanced accuracy. Its fused multiply-accumulate operations avoid intermediate rounding, ensuring exact numerical reproducibility through the quire, a wide fixed-point accumulator spanning the format's full dynamic range to prevent precision loss and overflow during long accumulations. However, integrating such large accumulators incurs significant area and power overheads. This paper presents an optimized, high-frequency \emph{Multispeculative PositMAC} architecture for 32- and 64-bits Posit. First, the pipeline is restructured to balance the different stages. Second, high-speed multiplication topologies are evaluated, showing that a \emph{Booth-4} scheme with \emph{Kogge--Stone} adders meets a stringent \delay{0.5} target (\frequency{2}). Finally, the wide monolithic quire accumulator is replaced with a \emph{Multispeculative Adder}, diminishing area up to $19.8\%$ while reducing energy consumption by more than $50\%$ when compared to the baseline. Compared to other state-of-the-art designs, our proposal achieves the highest operating frequency and reduces cycle time by up to $79.0\%$ with respect to 64-bit quire-enabled alternatives. This performance is attained without increasing resource overhead, as the design remains strictly smaller in area and achieves lower per-cycle energy consumption than all quire-capable counterparts.

\end{abstract}

}

\maketitle

\IEEEdisplaynontitleabstractindextext
\IEEEpeerreviewmaketitle

\begin{acronym}[]
    \acro{AI}{Artificial Intelligence}
    \acro{ALU}{Arithmetic Logic Unit}
    \acro{ArTeCS}{group of Architecture and Technology of Computing Systems}
    \acro{ASIC}{Application-Specific Integrated Circuit}
    \acro{BiCG}{Biconjugate Gradient}
    \acro{BLAS}{Basic Linear Algebra Subprogram}
    \acro{BSC}{Barcelona Supercomputing Center}
    \acro{CAPAP-H}{Red de Computación de Altas Prestaciones sobre Arquitecturas Paralelas Heterogéneas}
    \acro{CISC}{Complex Instruction Set cCmputer}
    \acro{CNN}{Convolutional Neural Network}
    \acro{CG}{Conjugate Gradient}
    \acro{RTU}{Reconfigurable Tensor Unit}
    \acro{DFMA}{Dynamic Fused MAC}
    \acro{DARE}{Digital Autonomy with RISC-V in Europe}
    \acro{DNN}{Deep Neural Network}
    \acro{DSL}{Domain Specific Language}
    \acro{DSP}{Digital Signal Processing}
    \acro{EPI}{European Processor Initiative}
    \acro{FF}{Flip-Flop}
    \acro{FFT}{Fast Fourier Transform}
    \acro{FMA}{Fused Multiply-Add}
    \acro{FPGA}{Field-Programmable Gate Array}
    \acro{FPU}{Floating-Point Unit}
    \acro{FU}{Functional Unit}
    \acro{GPU}{Graphics Processing Unit}
    \acro{GAN}{Generative Adversarial Network}
    \acro{GEMM}{general matrix multiplication}
    \acro{HPC}{High-Performance Computing}
    \acro{HLS}{High-Level Synthesis}
    \acro{HUB}{Half-Unit-Biased}
    \acro{IP}{Intellectual Property}
    \acro{IR}{Intermediate Representation}
    \acro{ISA}{Instruction Set Architecture}
    \acro{KPI}{Key Performance Indicator}
    \acro{LAPACK}{Linear Algebra Package}
    \acro{LLM}{Large Language Model}
    \acro{LSB}{Least Significant Bit}
    \acro{LUT}{Lookup Table}
    \acro{ML}{Machine Learning}
    \acro{MAC}{Multiply-Accumulate}
    \acro{MSE}{Mean Squared Error}
    \acro{MSB}{Most Significant Bit}
    \acro{MSADD}{Multispeculative Adder}
    \acro{MaxAbsE}{maximum absolute error}
    \acro{NaN}{Not a Number}
    \acro{NaR}{Not a Real}
    \acro{NN}{Neural Network}
    \acro{NoC}{Network on Chip}
    \acro{OS}{Operating System}
    \acro{OTRI}{Oficina de Transferencia de Resultados de Investigación}
    \acro{PAU}{Posit Arithmetic Unit}
    \acro{PNS}{Posit Number System}
    \acro{PM}{Person Month}
    \acro{PositMAC}{Posit MAC}
    \acro{PositMAC32}{32--bits Posit MAC}
    \acro{PositMAC64}{64--bits Posit MAC}
    \acro{Posit8}{8--bits Posit}
    \acro{Posit16}{16--bits Posit}
    \acro{Posit32}{32--bits Posit}
    \acro{Posit64}{32--bits Posit}    
    \acro{QC}{Quantum Computing}
    \acro{QtP}{Quire-to-Posit}
    \acro{RISC}{Reduced Instruction Set Computer}
    \acro{RTL}{Register-Transfer Level}
    \acro{RMSE}{Root Mean Squared Error}
    \acro{SGA}{Specific Grant Agreement}
    \acro{SIMD}{Single Instruction, Multiple Data}
    \acro{SNN}{Spiking Neural Networks}
    \acro{SOHA}{Spanish Open Hardware Alliance}
    \acro{SORN}{Sets Of Real Numbers}
    \acro{TALU}{Transprecision ALU}
    \acro{UCM}{Complutense University of Madrid}
    \acro{UPC}{Polytechnic University of Catalonia}
    \acro{VHDL}{VHSIC Hardware Description Language}
    \acro{VHSIC}{Very High Speed Integrated Circuit}
    \acro{WP}{Work Package}

\end{acronym}

\ifCLASSOPTIONcompsoc
\IEEEraisesectionheading{\section{Introduction}\label{sec:introduction}}
\else
\section{Introduction}
\label{sec:introduction}
\fi

\IEEEPARstart{T}{he} IEEE 754 standard is the most widely used representation of real numbers in computers~\cite{IEEE2019}. Although robust and reliable, it has several limitations: results can be inconsistent across platforms, addition and multiplication are not guaranteed to be associative, and it provides an excess of \ac{NaN} representations. Posit arithmetic was proposed by John Gustafson in 2017~\cite{Gustafson2017Beating} as an alternative to IEEE 754 for representing and operating with real numbers. Posits provide reproducible results across platforms and few special cases. They also avoid overflow and underflow and do not waste patterns on \acp{NaN}, reducing exception-handling complexity. Since their introduction, posits have received considerable attention~\cite{mallasen2025navigating,zhang2024review}, leading to RISC-V cores supporting posit arithmetic~\cite{mallasen2024BigPERCIVAL,sharma2023CLARINET,Tiwari2021,arunkumar2020PERC,cococcioni2022Lightweight} and even posit-based accelerators~\cite{prabhu2025minotaur,hao2025positCIM,murillo2023Generating,lu2021Evaluations,nakasato2024Evaluation,ramachandran2024AlgorithmHardware,ledoux2022Generator}.

Both IEEE 754 and posit arithmetic support fused multiply--accumulate (\ac{MAC}) operations. While IEEE 754 rounds each \ac{MAC}, the posit standard extends the \emph{fused} concept through a large register, the \emph{quire}, which stores accumulated results without intermediate rounding until the \ac{MAC} sequence is complete. However, implementing the quire and fused operations significantly increases area and power. As shown in \cite{mallasen2022PERCIVAL}, the quire register and associated operations account for approximately half the area of posit arithmetic units. Its size also introduces significant accumulation delay~\cite{murillo2021EnergyEfficient,sharma2023CLARINET,mallasen2024BigPERCIVAL}, limiting posit \ac{MAC} throughput. Beyond the quire, the multiplication stage dominates the unit's area, power, and delay, further limiting its target frequency.

Literature outlines two main quire accumulator approaches: single monolithic high-performance adders (like Kogge–Stone~\cite{kogge1973parallel}) that deliver low latency at the expense of high power and area, and multi-segment pipelined designs ~\cite{sharma2023CLARINET} that achieve higher clock frequencies but introduce severe latency penalties due to multi-stage carry propagation, especially in short sequences. As a way of mitigating this penalty, in this paper we utilize the \ac{MSADD}, introduced in \cite{delbarrio2012multispeculative} to improve the performance of large adders when accumulating sequences of additions. This design consists of several configurable-size blocks that operate independently, without propagating carry-out signals between blocks within the same cycle. Instead, these signals are stored and fed to the following block in the next cycle. This reduces both adder delay and area, as the area of large fast prefix adders scales as $O(n·log(n))$, where $n$ is the adder bit width. Finally, during the last addition of the sequence, the MSADD enters the \emph{speculation} phase, predicting the carry-in of each block to avoid propagating carries from the first to the last block, thereby achieving high performance.

All in all, we introduce novel 32-bit and 64-bit posit multiply--accumulate units based on the posit \ac{MAC} presented in \cite{murillo2021EnergyEfficient}. The proposed units restructure the pipeline, optimize the largest components, and incorporate the MSADD in the final accumulation stage, enabling higher operating frequencies than state-of-the-art designs while reducing area and energy consumption. Concretely, our designs are able to comply with a stringent 2 GHz frequency constraint, while diminishing 19.5\% and 17.6\% area and reducing energy by more than 50\% with respect to the baseline. When compared with state-of-the-art quire-based designs, our proposal also prevail in the three angles (frequency, area and energy).

The rest of this paper is organized as follows: Section~\ref{sec:posit arithmetic} introduces posit arithmetic, Section~\ref{sec:related work} presents related work and baseline design analysis, and Section~\ref{sec:mspMAC} details the proposed \emph{Multispeculative PositMAC} unit\footnote{Design URL:~\url{https://github.com/MarioInf-Phd-ComputerScience-UCM/MS_PositMAC}}. Section~\ref{sec:experiments} evaluates synthesis results, performance, and comparisons with state-of-the-art designs, while Section~\ref{sec:conclusions} concludes the paper.
\section{Background}
\label{sec:posit arithmetic}

\noindent
In this section, several relevant concepts will be introduced. These are essential to understand the rest of the paper. 

\subsection{Posit Arithmetic}
\label{subsec:posit arithmetic}

\noindent
A posit number is typically defined by two parameters, $\langle n, es \rangle$, where $n$ is the total bit width and $es$ is the number of exponent bits. While the most common formats in the literature~\cite{Gustafson2017Beating, DeDinechin2019, Murillo2020Deep} are \positenv{8}{0}, \positenv{16}{1}, and \positenv{32}{2}, the latest Posit Standard, published in 2022~\cite{positworkinggroup2022Standard}, fixes $es=2$. This simplifies hardware design and facilitates conversion between posit sizes~\cite{Guntoro2020}, at the cost of reduced dynamic range for $n>16$. Therefore, all posit configurations used in this paper are \positenv{n}{2}, denoted as Posit$n$.

Unlike IEEE 754 numbers, posit arithmetic defines only two special cases: zero and \ac{NaR}, represented by \texttt{0$\dots$0} and \texttt{10$\dots$0}, respectively. All other bit patterns represent real values according to the format in \figurename~\ref{fig:posit_format}:

\begin{figure}[!b]
    \centering
    \includegraphics[width=1.0\linewidth]{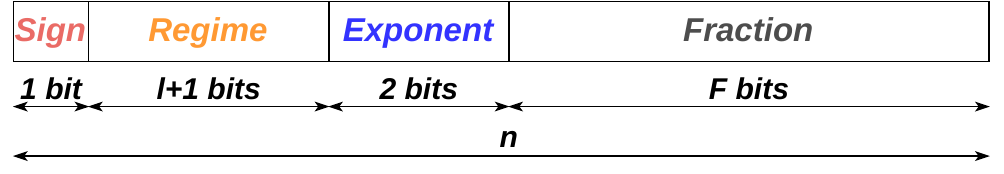}
    \caption{Generic $n$-bit posit format.}
    \label{fig:posit_format}
\end{figure}

\begin{itemize}
    \item \textbf{\emph{Sign}}: Sign bit.

    \item \textbf{\emph{Regime}}: Variable-length sequence of $l+1$ identical bits ($r$) terminated by its negation ($\bar{r}$), encoding the scaling factor $k$ given by Equation~\eqref{eq:regime_value}.
    
    \item \textbf{\emph{Exponent}}: Up to $es=2$ unbiased exponent bits. Most values use $es=2$ bits, while corner cases with $n-2$ or $n-1$ regime bits contain only one or zero exponent bits, respectively.

    \item \textbf{\emph{Fraction}}: Variable-length sequence representing the normalized fraction $f$ ($0 \leq f < 1$), calculated as the field's unsigned integer value divided by $2^F$, where $F$ ($0 \leq F \leq n-5$) is the number of fraction bits.
\end{itemize}

\begin{equation} \label{eq:regime_value}
    k = \left\{
    \begin{array}{ll}
        -l & \mbox{if } r_0 = 0 \\
        l-1 & \mbox{if } r_0 = 1
    \end{array}
    \right..
\end{equation}

Given a posit bitstring $P$,  Equation~\eqref{eq:posit_value} defines a generic posit's real value $X$. Unlike floating-point formats with a fixed hidden bit (1, or 0 for subnormals), posits interpret this bit as 1 if positive or $-2$ if negative. Although two's complement decoding style has proved to be faster than the sign and magnitude interpretation, it increases fixed-point operand widths in multiplicative designs~\cite{Guntoro2020, murillo2022Comparing}. Thus, in this paper we adopt the sign and magnitude approach shown in Equation~\eqref{eq:posit_value}, the same as in~\cite{murillo2021EnergyEfficient}. It must be noted that if $P<0$, then it must be negated prior to being decoded.

\begin{equation} \label{eq:posit_value}
    X = (-1)^{\textcolor{red}{s}}
    \cdot
    2^{\textcolor{orange}{k}\,2^{es} + \textcolor{blue}{e}}
    \cdot
    \left(1 + \textcolor{gray}{f}\right)~.
\end{equation}

\begin{table} [!b]
\centering
\ra{1.2}
\caption{Quire bitwidth parameters for different posit formats. Note that only formats with $es=2$ are standard.}
\label{tab:quireBitwidthParameters}
    \begin{tabular}{lcccc}
    \toprule
    \multicolumn{2}{c}{Posit}    &
    \multicolumn{3}{c}{Quire}    \\ 
    \cmidrule(lr){1-2}
    \cmidrule(lr){3-5}
    
    $n$ &
    $es$ &
    $C$ &
    $maxE$ &
    $qSize$ \\
    \midrule
    
    8 & 0 & 7 & 6 & 32 \\
    8 & 1 & 15 & 12 & 64 \\
    8 & 2 & 31 & 24 & 128 \\
    16 & 1 & 15 & 28 & 128 \\
    16 & 2 & 31 & 56 & 256 \\
    32 & 2 & 31 & 120 & 512 \\
    64 & 2 & 31 & 248 & 1024 \\
    \bottomrule
    \end{tabular}
\end{table}

\begin{table*}[!b]
    \centering
    \caption{State-of-the-art designs supporting multiply–accumulate operations using the posit number format.}
    \label{table:Other_hardware_implementations}
    
    \rowcolors{2}{gray!15}{white}
    
    \resizebox{\textwidth}{!}{
        \begin{tabular}{cccccccc}
            \toprule Design     
                & Data size                 & Parameterized     & Pipeline          & Quire         & Implementation    & Technology        & Open source \\
            \midrule
            
            Carmichael \etal~\cite{Carmichael2019a} (Deep Positron)
                & Posit8                        & $\checkmark^\ast$ & $\checkmark$      & $\checkmark$  & VHDL           & Virtex-7          & $\checkmark$ \\
            
            Zhang \etal~\cite{Zhang2019}
                & Posit8 / Posit16 / Posit32    & $\checkmark$      & $\checkmark$      & $\times$      & Verilog           & ASIC              & $\times$ \\
            
            Uguen \etal~\cite{Uguen2019} (MArTo)
                & Posit16 / Posit32             & $\checkmark$      & $\checkmark$      & $\checkmark$  & Vitis HLS         & Kintex-7          & $\checkmark$ \\
            
            Neves \etal~\cite{Neves2020}
                & Posit8 / Posit16 / Posit32    & $\checkmark$      & $\checkmark$      & $\checkmark$  & ---               & Virtex-7 / ASIC   & $\times$ \\

            Crespo \etal~\cite{crespo2023trading}
                & Posit8 / Posit16              & $\checkmark$      & $\checkmark$      & $\checkmark$  & VHDL              & ASIC              & $\checkmark$ \\
            
            Murillo \etal~\cite{murillo2021EnergyEfficient} (Baseline design)
                & Posit8 / Posit16 / Posit32    & $\checkmark$      & $\checkmark$      & $\checkmark$  & VHDL$^\dagger$    & ASIC              & $\checkmark$ \\
            
            Li \etal~\cite{li2023PDPU}
                & Posit16                   & $\checkmark$      & $\checkmark$      & $\checkmark$  & SystemVerilog     & ASIC              & $\checkmark$ \\
            
            Kumar \etal~\cite{kumar2026spade}
                & Posit8 / Posit16 / Posit32    & $\checkmark$      & $\checkmark$      & $\checkmark$  & Verilog           & Virtex-7 / ASIC   & $\times$\\
            
            \bottomrule
        \end{tabular}
    }
    
    \vspace{0.5em}
    
    {\footnotesize
    $\checkmark^\ast$ Partially supported.
    $^\dagger$ Generated with FloPoCo~\cite{dedinechin2011Designing}.}
    
\end{table*}

\subsection{Fused operations}
\label{subsec:fused operations}

\noindent
Since the 2008 IEEE 754 revision, floating-point arithmetic has included a fused multiply-add operation~\cite{IEEEComputerSociety2008}. In contrast, the posit standard~\cite{positworkinggroup2022Standard} has included fused operations from its inception to perform consecutive \acp{MAC}. However, \emph{fused} has a broader meaning for posits than for IEEE 754, as values are accumulated in a large register called \emph{quire}, avoiding intermediate rounding until the \ac{MAC} sequence is complete. This provides higher accuracy and reproducibility~\cite{murillo2021EnergyEfficient, crespo2022Unified, mallasen2024BigPERCIVAL}. This idea resembles the Kulisch accumulator used in floating-point arithmetic~\cite{kulisch2008Computer}. However, the Kulisch accumulator is not part of IEEE 754 and is slightly larger than the quire.

The precision of the quire format is determined by the dynamic range of the corresponding \positenv{n}{es} format. The largest and smallest positive posit values are $2^{maxE}$ and $2^{-maxE}$, respectively, where $maxE$ is largest exponent for such a format, as defined by Equation \ref{eq:max_exp}.

\begin{equation} \label{eq:max_exp}
    maxE = (n-2)\times 2^{es}~.
\end{equation}

\begin{figure}[!b]
    \centering
    \includegraphics[width=1\linewidth]{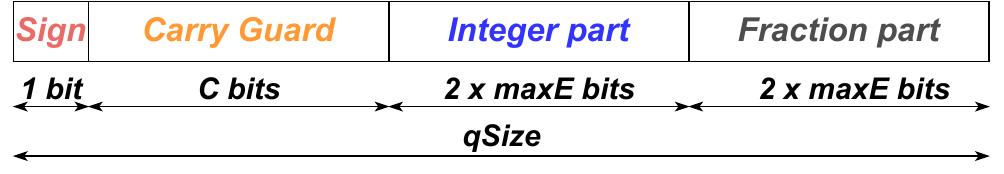}
    \caption{quire format encoding for \positenv{n}{es}. 
    }
    \label{fig:quire_format}
\end{figure}

To correctly represent $2^{maxE}\times2^{maxE}$ in fixed-point format we need $2 \times maxE$ integer bits, plus one more \emph{carry bit} in the quire. To correctly represent $2^{-maxE}\times2^{-maxE}$ we need $2 \times maxE$ fractional bits plus one integer bit, so that the quire is organized as depicted in \figurename~\ref{fig:quire_format}:

\begin{itemize}
    \item\textbf{\emph{Sign}}: 1 sign bit.
    
    \item\textbf{\emph{Integer}}: $2 \times maxE$ bits to represent the integer part.
    
    \item\textbf{\emph{Fraction}}: $2 \times maxE$ bits for the fractional part.

    \item\textbf{\emph{Carry Guard}}: $C$ bits allowing up to $2^{C}-1$ product sums without overflow. A common rule of thumb is to choose $C$ so that the total quire length ($qSize$), including the sign bit, is a power of two. The latest posit standard specifies $es=2$, giving $qSize=16n$ bits and a fixed 31 carry bits.
\end{itemize}

\section{Related Work}
\label{sec:related work}

\noindent
Since the introduction of posits, various hardware designs have been proposed to compare posit arithmetic against their IEEE~754 floating-point counterparts. For instance, Uguen et al.~\cite{Uguen2019} presented a customizable posit C++ library for Vivado HLS covering addition, subtraction, and multiplication. The reported results showed floating-point units achieving lower latency and resource usage than posit units.

Subsequent work has further explored specialized posit hardware architectures. Neves et al.~\cite{neves2021Reconfigurable} proposed a \ac{RTU} based on an array of variable-precision posit vector MAC units. Similarly, \cite{wu2025pvu} presents a Posit Vector Arithmetic Unit, while \cite{li2023PDPU} introduces a six-stage pipelined Posit Dot Product Unit. Additionally; Dube et al.~\cite{dube2025compact} presented a \ac{TALU} supporting multiple formats, including posits; while Condia et al.~\cite{condia2025investigating} evaluated permanent hardware fault effects in posit versus floating-point units.

Various posit \ac{MAC} designs have targeted higher performance and lower area. Neves et al.~\cite{Neves2020} proposed a \ac{DFMA} supporting variable exponent sizes, while Crespo et al.~\cite{crespo2022Unified} presented a unified Posit and IEEE-754 vector MAC execution to reduce area and power consumption. Murillo et al.~\cite{murillo2021EnergyEfficient} introduced a scalable posit MAC architecture designed independently of the overall posit bit width and evaluated across \ac{Posit8}, \ac{Posit16}, and \ac{Posit32} operations, which serves as our baseline design for the present work.

Additionally, Crespo et al.~\cite{crespo2023trading} explored area and power savings via low-precision posit \ac{MAC} units, while Crafton et al.~\cite{crafton2025finding} evaluated the cost, power, and numerical accuracy trade-offs for posits and other emerging formats. Posit support has also been integrated into specialized processing cores. Mallasén et al.~\cite{mallasen2022PERCIVAL} introduced \emph{PERCIVAL}, the first RISC-V core with a full posit instruction set, incorporating the \ac{Posit32} MAC unit from~\cite{murillo2021EnergyEfficient}. They later expanded this architecture in \emph{Big-PERCIVAL}~\cite{mallasen2024BigPERCIVAL} to support 64-bit posit operations.

Other general-purpose cores have also been proposed. Sharma et al.~\cite{sharma2023CLARINET} presented \emph{Melodica}, a posit unit with fused operations integrated into the \emph{Clarinet} RISC-V core. Similarly, Mallasén et al.~\cite{mallasen2025phee} introduced \emph{PHEE}, a modular posit coprocessor integrated into the X-HEEP RISC-V SoC~\cite{machetti2024xheep}. Brownfield extensions supporting posit units were also presented in~\cite{arunkumar2020PERC,Tiwari2021}. Table~\ref{table:Other_hardware_implementations} summarizes state-of-the-art hardware implementations supporting posit multiply--accumulate operations besides key implementation details.

\subsection{Baseline PositMAC unit architecture}
\label{subsec:Baseline Posit MAC Unit architecture}

\begin{table*}[!b]
    \centering
    \caption{Baseline PositMACs synthesis results obtained using Synopsys Design Compiler with 28\,nm, TSMC standard-cell library and \delay{0.5} timing constraint and preserving the internal component hierarchy.}
    \label{table:PositMAC_InitialValues_0.5ns}

    \fontsize{8.5pt}{8pt}\selectfont      
    \setlength{\tabcolsep}{6pt}         

    \def\units{false}
    \def\headerUnits{true}
    
    \begin{tabular}{l!{\vrule width 1pt} c c c c c c !{\vrule width 1pt} c c c c c c c}
        \toprule
        &   \multicolumn{6}{c}{\rule{0pt}{1.4em}\textbf{\textit{\normalsize Baseline PositMAC 32}} } {\vrule width 1pt}&
            \multicolumn{7}{c}{\rule{0pt}{1.4em}\textbf{\textit{\normalsize Baseline PositMAC 64}}} \\[4pt]
        
        &   \textbf{Area} & \textbf{Power} & \textbf{Energy} & \multicolumn{3}{c}{\textbf{Delay}} {\vrule width 1pt}&
            \textbf{Area} & \textbf{Power} & \textbf{Energy} & \multicolumn{4}{c}{\textbf{Delay}} \\

        &   (\area{}[\headerUnits]) & (\power{}[\headerUnits]) & (\energy{}[\headerUnits]) & \multicolumn{3}{c}{(\delay{}[\headerUnits])} {\vrule width 1pt}&
            (\area{}[\headerUnits]) & (\power{}[\headerUnits]) & (\energy{}[\headerUnits]) & \multicolumn{4}{c}{(\delay{}[\headerUnits])} \\

        \cmidrule(lr){5-7}
        \cmidrule(lr){11-14}
        
        & & & & \textbf{St.1} & \textbf{St.2} & \textbf{St.3} &
        & & & \textbf{St.1} & \textbf{St.2} & \textbf{St.3} & \textbf{St.4} \\
        \midrule

        \textbf{\textit{Decode A}}          & 
        \area{428.40}[\units]          & 
        \power{0.16}[\units] & 
        \energy{0.22}[\units]    & 
        \delay{0.27}[\units] &
        -- &
        -- &
        
        \area{656.84}[\units]          & 
        \power{0.35}[\units]    & 
        \energy{0.55}[\units]        &
        --          &
        -- &
        -- &
        -- \\
        
        \textbf{\textit{Decode B}}          & 
        \area{533.11}[\units]          & 
        \power{0.18}[\units]    & 
        \energy{0.23}[\units]   &
        -- &
        -- & 
        -- &
        
        \area{690.86}[\units]          & 
        \power{0.21}[\units]    & 
        \energy{0.33}[\units] &
        \delay{0.40}[\units] & 
        -- & 
        -- & 
        -- \\


        \textbf{\textit{Multiplier}}        & 
        \area{6065.14}[\units]          & 
        \power{6.32}[\units]     &
        \energy{8.17}[\units]    &
        \delay{0.80}[\units]          &
        -- & 
        -- &
        
        \area{18571.52}[\units]        & 
        \power{18.78}[\units]   & 
        \energy{29.30}[\units]  &
        \delay{0.96}[\units]          & 
        -- & 
        -- & 
        -- \\

        \textbf{\textit{Frac. Normalizer}}  &
        \area{69.05}[\units]          & 
        \power{0.11}[\units]    & 
        \energy{0.14}[\units]   &
        \delay{0.01}[\units]          &
        -- & 
        -- &
        
        \area{148.30}[\units]          & 
        \power{0.23}[\units]    & 
        \energy{0.36}[\units]  &
        \delay{0.01}[\units]          & 
        -- & 
        -- & 
        -- \\


        \textbf{\textit{SF Adder}}          & 
        \area{92.86}[\units]          & 
        \power{0.08}[\units]    & 
        \energy{0.10}[\units]  &
        \delay{0.03}[\units]          & 
        -- & 
        -- &
        
        \area{98.91}[\units]           & 
        \power{0.08}[\units]    & 
        \energy{0.13}[\units]        &
        \delay{0.04}[\units]          & 
        --          & 
        -- & 
        -- \\
        
        \textbf{\textit{Rounding Adder}}    & 
        \area{95.00}[\units]          & 
        \power{0.06}[\units]    & 
        \energy{0.07}[\units]        &
        \delay{0.05}[\units]          &
        -- & 
        -- &
        
        \area{118.31}[\units]           & 
        \power{0.07}[\units]    & 
        \energy{0.11}[\units]        &
        \delay{0.05}[\units]          & 
        -- & 
        -- & 
        -- \\


        \textbf{\textit{SF Biased Adder}}   & 
        \area{67.79}[\units]          & 
        \power{0.053}[\units]    & 
        \energy{0.07}[\units]        &
        \delay{0.06}[\units]          & 
        -- & 
        -- &
        
        \area{99.79}[\units]           & 
        \power{0.06}[\units]    & 
        \energy{0.09}[\units]        &
        \delay{0.03}[\units] &
        -- & 
        -- & 
        -- \\

        \textbf{\textit{Shifter}}           & 
        \area{3221.69}[\units]         & 
        \power{2.31}[\units]    & 
        \energy{2.99}[\units]        &
        \delay{0.06}[\units] & 
        \delay{0.35}[\units] & 
        -- & 
        
        \area{8220.74}[\units]         & 
        \power{7.37}[\units]    & 
        \energy{11.49}[\units]        &
        \delay{0.04}[\units]          & 
        \delay{0.33}[\units]          & 
        \delay{0.28}[\units]          & 
        \delay{0.22}[\units]          \\

        \textbf{\textit{Quire Adder}}       & 
        \area{5950.35}[\units]          & 
        \power{4.66}[\units]     &
        \energy{6.02}[\units]   &
        -- & 
        -- & 
        \delay{0.39}[\units]          &

        \area{16504.99}[\units]         & 
        \power{11.41}[\units]     & 
        \energy{17.80}[\units]   &
        -- & 
        -- & 
        -- & 
        \delay{0.34}[\units]          \\
        
        \midrule
        \textbf{\textit{TOTAL}}             & 
        \area{20591.424}[\units]         & 
        \power{19.264}[\units]    & 
        \energy{24.91}[\units]   &
        \delay{1.29}[\units]          & 
        \delay{0.38}[\units]          & 
        \delay{0.50}[\units]          &
        
        \area{52949.36}[\units]         & 
        \power{47.03}[\units]    & 
        \energy{73.37}[\units]  &
        \delay{1.56}[\units]          & 
        \delay{0.34}[\units]          & 
        \delay{0.30}[\units]          & 
        \delay{0.61}[\units]          \\

        \bottomrule
        
    \end{tabular}
\end{table*}

\noindent
Authors in~\cite{murillo2021EnergyEfficient} present a posit \ac{MAC} architecture, which serves as our baseline. It multiplies two posit inputs (\emph{A} and \emph{B}) and adds the product to the previous accumulation in fixed-point format (\emph{C}). To prevent precision loss from intermediate roundings, it uses a quire register, albeit at the cost of higher latency and energy consumption than previous designs~\cite{Carmichael2019a,Uguen2019}. The execution steps of the baseline posit MAC unit are detailed below:

\begin{enumerate}    
    \item\textbf{\emph{Decoding posit operands}}: Extracts fields from inputs \emph{A} and \emph{B}, including the fraction \emph{frac} and the scaling factor \emph{sf}, which is determined by the regime and exponent bits, if any. In addition, the Zero and \ac{NaR} flags are determined during this stage. 
    
    \item\textbf{\emph{Multiplication of A and B}}: Computes the product of fractions \emph{frac}, accumulation of the scaling factors \emph{sf}, and normalizes the resulting value.

    \item\textbf{\emph{Converting into fixed-point format}}: Shifts the normalized product by scaling factors \emph{sf} into fixed-point format, using the bit widths listed in Table~\ref{tab:quireBitwidthParameters}.

    \item\textbf{\emph{Fixed-point accumulation}}: Adds the fixed-point product to input \emph{C}, which stores the accumulated result from previous operations.

\end{enumerate}

Additionally, posit MAC units require two external components: a \emph{qSize}-bit register (\emph{Quire Register}), whose size is shown in Table~\ref{tab:quireBitwidthParameters}, to store the fixed-point accumulation result, and a dedicated quire-to-posit converter. the works in \cite{mallasen2022PERCIVAL,sharma2023CLARINET} present the hardware architecture of the \ac{PAU}, which integrate these external components along with MAC, logical, and arithmetic posit units. According to the posit standard, the posit MAC unit supports the following quire operations:

\begin{itemize}
    \item \textbf{\texttt{qMulAdd(quire, posit1, posit2):}} Returns the result of adding the product of \emph{posit1} and \emph{posit2} to \emph{quire}, i.e., \emph{quire} + (\emph{posit1} $\times$ \emph{posit2}).

    \item \textbf{\texttt{qMulSub(quire, posit1, posit2):}} Returns the result of subtracting the product of \emph{posit1} and \emph{posit2} from \emph{quire}, i.e., \emph{quire} $-$ (\emph{posit1} $\times$ \emph{posit2}).

    \item \textbf{\texttt{qNegate(quire):}} Returns the negated value of \emph{quire}, i.e., $-\emph{quire}$.
\end{itemize}

\begin{figure}[t]
    \centering
    \subfloat[PositMAC32.\label{fig:PositMAC32_architecture}]{%
        \includegraphics[width=0.49\linewidth]{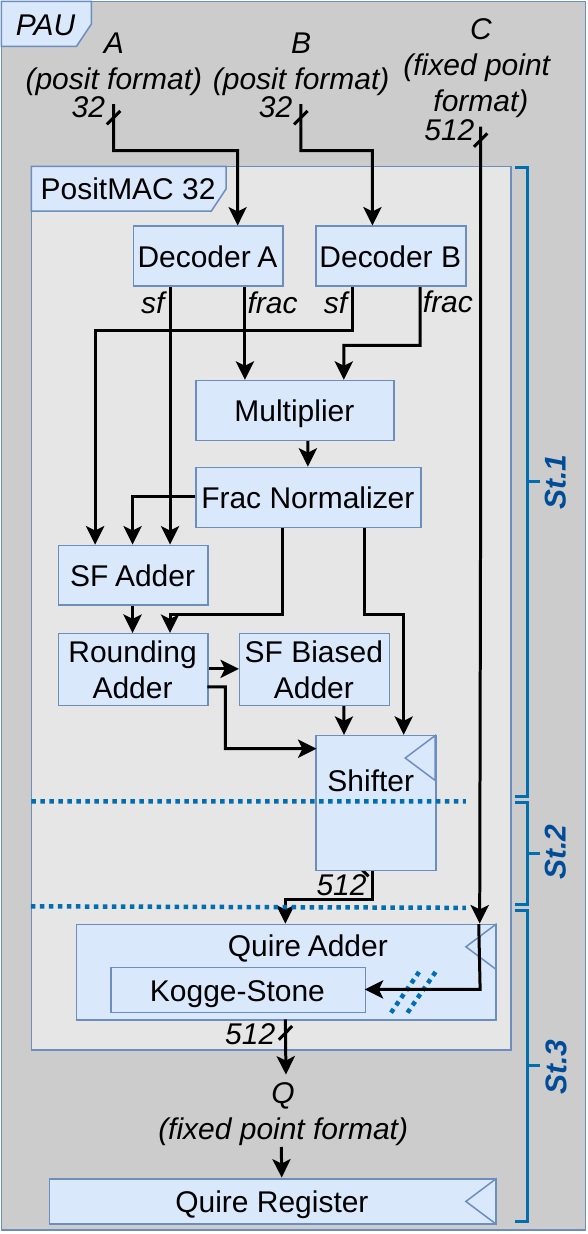}%
    }
    \hfill
    \subfloat[PositMAC64.\label{fig:PositMAC64_architecture}]{%
        \includegraphics[width=0.49\linewidth]{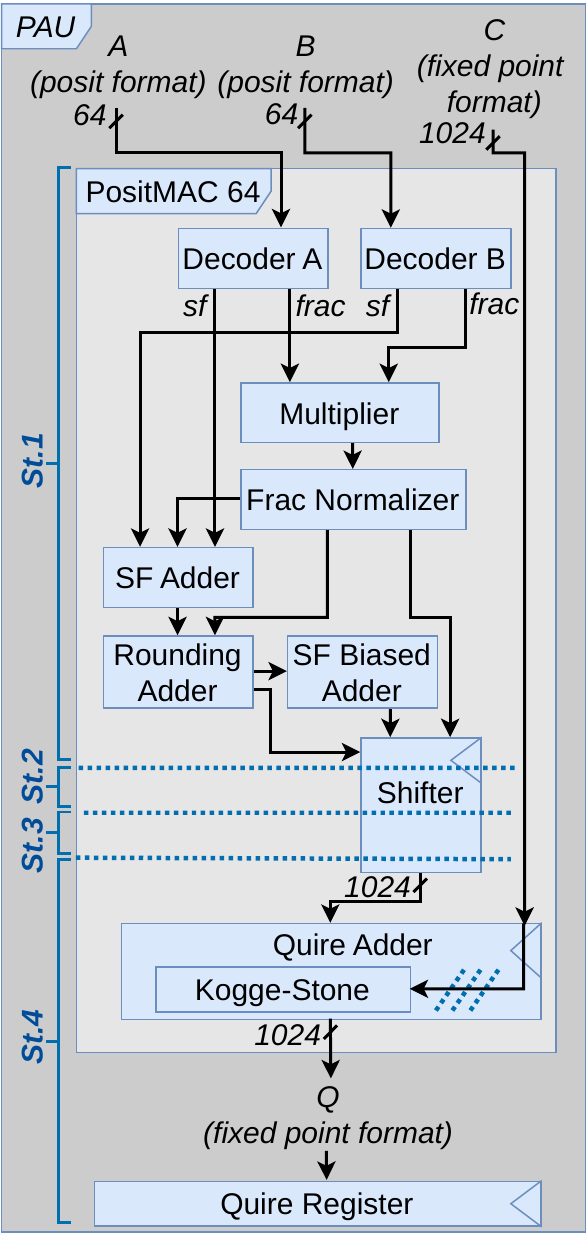}%
    }
    \caption{Architectures of the baseline posit MAC units.}
    \label{fig:PositMAC_architecture}
\end{figure}

In addition, the unit implements a custom \emph{\texttt{QClear}} operation to reset its internal state and quire register. The top-level unit, integrating both \ac{Posit32} and \ac{Posit64} \ac{MAC} implementations, handles all control signals and input parameters required for these operations. The baseline \ac{PositMAC} architecture supports pipelining, with registers inserted automatically via FloPoCo~\cite{dedinechin2011Designing,Murillo2020Customized}. As shown in Fig.~\ref{fig:PositMAC_architecture}, the \ac{Posit32} and \ac{Posit64} variants feature three and four pipeline stages, respectively. Neither unit includes dedicated logic for inter-stage data transition management.

\subsection{Baseline PositMAC unit synthesis}
\label{subsec:Baseline Posit MAC Unit synthesis}

\noindent
This section presents a preliminary evaluation of the baseline design~\cite{murillo2021EnergyEfficient}. Related \ac{Posit32} implementations synthesized in 28\,nm technology achieved maximum frequencies of 2.7~GHz~\cite{Zhang2019}, 0.62~GHz~\cite{Neves2020}, and 1.3~GHz~\cite{kumar2026spade}. It must be noted that the ~\cite{Zhang2019} is not a posit MAC unit, but a dot product unit for 16-bit operands, and the quired version frequency is not reported. Based on these results, we targeted a 2~GHz operating frequency. Synthesis was performed using a 28\,nm TSMC standard-cell library with a \delay{0.5} clock period and timing constraint. Constraints also specified a maximum fan-out of 200 pins and an assumed 0.1 toggle rate for non-clock registers and input signals.

Table~\ref{table:PositMAC_InitialValues_0.5ns} presents the synthesis results. Due to timing violations, energy consumption is computed as power multiplied by the critical-stage delay for each unit. These initial results highlight key issues addressed in this work:

\begin{itemize}

    \item The MAC architecture suffers from imbalanced pipeline stages, saturating the first stage while underutilizing others. This stems mainly from FloPoCo's optimization for \ac{FPGA} generation, yielding sub-optimal pipeline partitioning for \ac{ASIC} designs. Additionally, concentrating control registers inside the \emph{Shifter} increasing its area, power consumption, and timing delay.

    \item The \emph{Multiplier} component dominates area and power consumption while taking \delay{0.80} and \delay{0.96} to complete its operation in the \ac{Posit32} and \ac{Posit64} implementations, respectively, causing the first stage's total delay to exceed the \delay{0.5} target clock period.

    \item The \emph{Quire Adder} component's large input width yields high area and power consumption, which are further inflated by the required internal synchronization registers.
    
\end{itemize}

\section[Multispeculative Posit MAC]{Multispeculative Posit \ac{MAC}}
\label{sec:mspMAC}

\noindent
This section explains the process followed to develop the Multispeculative Posit MAC units, including both the Posit32 and Posit64 versions, starting from their respective baseline designs.

\subsection{Restructuring the pipeline}
\label{subsec:Restructuring the pipeline}

\begin{figure*}[!t]
    \centering
    \includegraphics[width=1.0\linewidth]{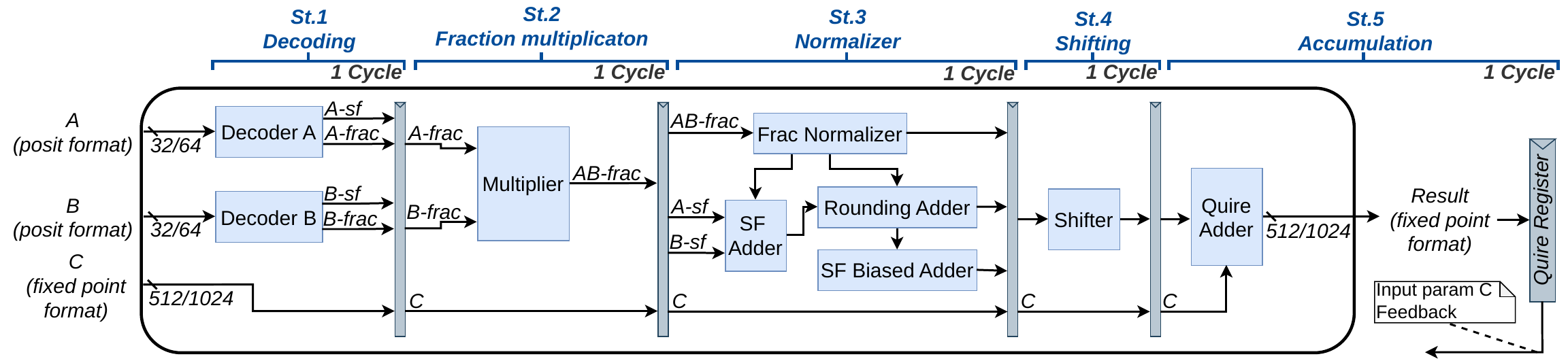}
    \vspace{-0.5em}
    \caption{\ac{PositMAC32}/\ac{PositMAC64} restructured pipeline.}
    \label{fig:PositMAC_restructured_pipeline}
\end{figure*}

\noindent
These preliminary results reveal an important issue related to stage delay balancing. In both units, the first stage, which comprises the subcomponents from \emph{Decoder A} to \emph{Shifter}, exhibits timing delays of \delay{1.21} and \delay{1.33} for the \ac{PositMAC32} and \ac{PositMAC64}, respectively. It must also be noted that the maximum delay among the remaining stages is only \delay{0.58} for the \ac{PositMAC32} and \delay{0.57} for the \ac{PositMAC64}. Due to the delay imbalance among the different stages, the results obtained for stages with lower delays may be misleading. This occurs because Synopsys Design Compiler concentrates most of the optimization effort on satisfying the timing constraints of the critical path, while applying a lower level of optimization to stages with substantially smaller delays.

In order to 1) correct the delay imbalance among stages; and 2) obtain more optimized results in the non-critical stages, the pipeline of the baseline unit has been restructured by separating the first-stage components with larger delay. In particular, the \emph{Multiplier} exhibits the highest delay, with \delay{0.80} and \delay{0.73} for the \ac{PositMAC64} and \ac{PositMAC32}, respectively. The resulting architecture shares the same pipeline structure for both \ac{PositMAC32} and \ac{PositMAC64} and is shown in Fig.~\ref{fig:PositMAC_restructured_pipeline}. It is composed of the following stages:

\begin{itemize}
    \item\textbf{Stage 1 - Decoding:} Extraction of the relevant fields from the input posits, including regime, exponent, and fraction, as well as the determination of the Zero and \acs{NaR} flags. The corresponding operations are performed by the \emph{Decoder A} and \emph{Decoder B} components.

    \item\textbf{Stage 2 - Fraction multiplication:} Computation of the product of the operand fractions, carried out by the \emph{Multiplier} component.

    \item\textbf{Stage 3 - Normalization:} Normalization of the intermediate result and adjustment of the associated scaling factors. This stage includes the \emph{Frac Normalizer}, \emph{SF Adder}, \emph{Rounding Adder}, and \emph{SF Biased Adder} components.

    \item\textbf{Stage 4 - Shifting:} Conversion of the normalized value into fixed-point representation through shifting according to the computed scale, performed by the \emph{Shifter} component.

    \item\textbf{Stage 5 - Accumulation:} Addition of the fixed-point value to the quire, which stores the accumulated result of previous operations in fixed-point format. This operation is handled by the \emph{Quire Adder} component.
\end{itemize}

\subsection{Optimizing the multiplier}
\label{subsec:Optimizing the multiplier}

\noindent
The synthesis results presented in Section \ref{subsec:Baseline Posit MAC Unit synthesis} reveal that the multiplier is the most demanding component of the \ac{PositMAC} architecture in terms of both timing and hardware resources. As shown in Table \ref{table:PositMAC_InitialValues_0.5ns}, it exhibits the highest propagation delay among all the arithmetic units, becoming the critical component of Stage~2 after restructuring the pipeline. Consequently, improving the multiplier is essential to further increase the maximum operating frequency of the proposed \ac{PositMAC} architecture.

This work evaluates two optimized multiplier implementations based on modified \emph{Booth} encoding~\cite{booth1951signed}, namely: radix-4 and radix-8 variations, which are widely accepted as the most efficient by literature~\cite{Ercegovac2004,delbarrio2019combined}, even though radix-8 needs to generate the $3X$ hard multiple. These are coupled with a 4:2 compressors reduction tree together with a \emph{Kogge--Stone} adder in the final addition stage of the multiplier.

\subsection{The Multispeculative Adder}
\label{subsec:The Multispeculative Adder}

\begin{figure}[!b]
    \centering
    \includegraphics[width=1.0\linewidth]{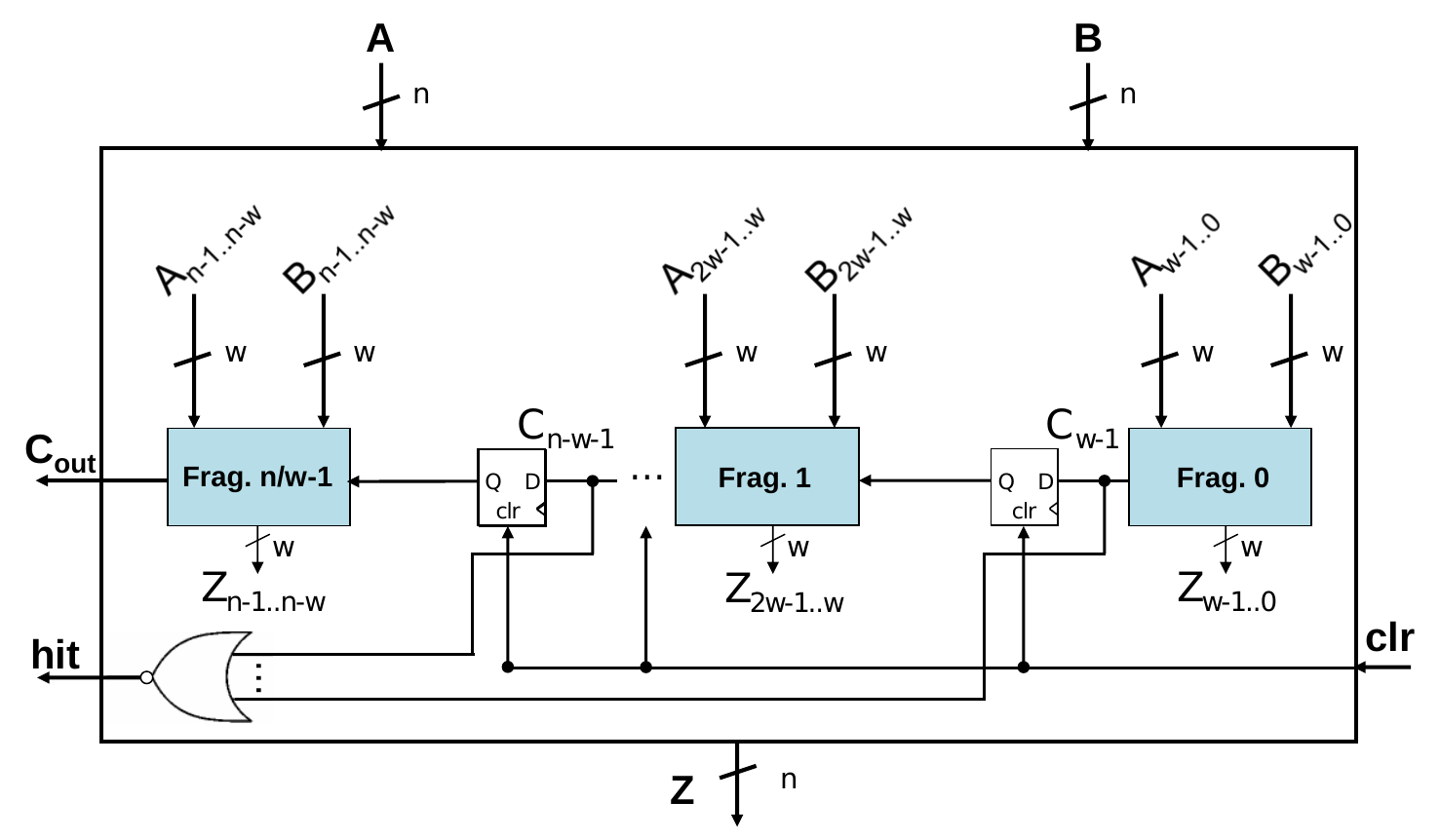}  
    
    \vspace{-1em}
    \caption{\ac{MSADD} with static zero prediction.}
    \label{fig:msadd_new}
\end{figure}

The accumulation stage is one of the main limiting factors in fused posit arithmetic. As shown in the synthesis results presented in Section \ref{subsec:Baseline Posit MAC Unit synthesis}, the \emph{Quire Adder} operates on very wide operands (512 bits for \ac{Posit32} and 1024 bits for \ac{Posit64}), which results in a significant contribution to the overall area, power consumption and timing delay of the MAC unit. Consequently, improving this stage represents an effective way to increase operating frequency while reducing implementation cost.

Quire accumulators in literature have followed two different approaches. The first relies on a monolithic high-performance adder, such as the \emph{Kogge--Stone} implementation adopted in the baseline \ac{PositMAC} architecture. This solution minimizes the latency of each accumulation but produces a considerable increase in hardware resources and power. The second approach divides the quire into several pipelined segments, as proposed in \emph{Clarinet/Melodica} \cite{sharma2023CLARINET}. Although this solution allows higher clock frequencies, carry propagation must traverse several pipeline stages, introducing a noticeable latency penalty that becomes particularly significant for short accumulation sequences.

The \ac{MSADD} overcomes these limitations by dividing the $N_q$-bit quire adder into $N_q/w$ independent fragments of width $w$, where $N_q$ denotes the quire width and each fragment can be implemented using any conventional adder architecture. Instead of propagating carries across all fragments during every accumulation, each fragment computes its local result independently while the generated carry is stored and forwarded to the following fragment during the next accumulation cycle. As a result, the critical path is determined by the delay of a single fragment instead of the complete quire adder. 

The original \ac{MSADD} architecture \cite{delbarrio2012multispeculative, delbarrio2016partial, delbarrio2019combined} was conceived as a general-purpose arithmetic unit capable of exchanging carry information with other multispeculative functional units through dedicated carry-in and carry-out interfaces. As a result, each fragment incorporates additional storage and control logic to receive carry signals generated by neighbouring \acp{MSADD} and to forward newly generated carries to subsequent units. In the context of fused posit arithmetic, the accumulation stage contains only one \ac{MSADD}, which is responsible for both generating and consuming all carry information internally. Therefore, the inter-unit carry communication required in previous \ac{MSADD}-based datapaths is no longer necessary.

Since only one \ac{MSADD} is required, the architecture can be simplified as shown in Fig. \ref{fig:msadd_new}. Rather than storing and forwarding carry information between different multispeculative units, only the carry-out bit generated by each fragment must be preserved from one accumulation cycle to the next. For this reason, a single D flip-flop per fragment is sufficient to store the carry-out signal before it is forwarded to the adjacent fragment during the following accumulation cycle. Given a sequence of $L$ additions, during the first $(L-1)$ accumulation cycles, the carry generated by fragment $(i)$ of one addition is propagated to fragment $(i+1)$ of the next addition. Once the final accumulation has been performed, the MSADD enters the \emph{speculation phase} assuming \emph{static zero prediction}. Thus, the remaining carry signals continue propagating between adjacent fragments until all carry-out values become zero and a hit is produced. This simplified architecture preserves the original multispeculative behaviour while reducing the control logic, hardware overhead and routing complexity.

The performance of the \ac{MSADD} can be analysed by modelling the probability of obtaining a correct speculative result, referred to as a hit, after each speculation cycle. Let $n$ denote the total bit width of the \ac{MSADD} and $w$ the width of each internal adder fragment. Thus, the architecture contains $n/w$ fragments, assuming that $n$ is an integer multiple of $w$. In the proposed \ac{PositMAC} units, $n$ corresponds to the quire width and is equal to 512 bits for \ac{PositMAC32} and 1024 bits for \ac{PositMAC64}. The average number of cycles required to complete a single multispeculative addition, denoted by $\lambda^1_{AVG}$, is given by Equation \ref{gather:avg_latency}. In this equation, the summation index $i$ identifies the speculation cycle under consideration, $P_{HIT}(i)$ represents the corresponding hit probability, and the product term accounts for the absence of a hit in the preceding cycles. Finally, note that in a conventional non-speculative case, $\lambda^1_{AVG}=1$ cycle.
\begin{gather}
\label{gather:avg_latency}
\lambda_{AVG}^{1}=\sum_{i=1}^{n/w}i*P_{HIT(i)}*\prod_{j=1}^{i-1}(1-P_{HIT(j)})~,\\
\label{gather:phit}
P_{HIT(\lambda)}=\prod_{i=\lambda}^{n/w-1}(1-P(M_{i-\lambda+2}^{1})*\prod_{j=0}^{\lambda-2}P(W_{i-j}))~.
\end{gather}
In Equation \ref{gather:phit}, $\lambda$ denotes the number of elapsed speculation cycles for which the hit probability is evaluated, $P(M^t_i)$ is the probability of a carry misprediction in fragment $i$ during cycle $t$, and $P(W_i)$ is the probability of finding a carry-propagate condition within fragment $i$.

Assuming equiprobable input data, the hit probability can be simplified to Equation \ref{gather:phit_equiprob}.
\begin{gather}
\label{gather:phit_equiprob}
P_{HIT(\lambda)}=\prod_{i=\lambda}^{n/w-1}\left(1-\frac{1}{2^{\lambda}}\right) = \left(1-\frac{1}{2^{\lambda}}\right)^{n/w-\lambda} ~.
\end{gather}

As can be observed, the maximum number of speculation cycles is limited to $n/w$. Therefore, the worst-case penalty is bounded by $n/w -1$ additional cycles. In contrast, increasing the fragment size improves the hit probability, reducing the average latency, whereas smaller fragments provide shorter critical paths and lower hardware cost.

When the \ac{MSADD} is employed in a sequence of fused MAC operations, the previous analysis can be extended to a sequence composed of $L$ consecutive accumulations. In this case, the average execution latency is given by Equation \ref{gather:avg_latency_sequence}.
\begin{gather}
\label{gather:avg_latency_sequence}
\lambda_{AVG}^{L}=L + \lambda_{AVG}^{1} - 1~.
\end{gather}

where $1 \leq \lambda_{AVG}^{1} \leq 2$ in many cases because data usually have temporal locality~\cite{delbarrio2012multispeculative}, $\lambda_{AVG}^{1} \in \mathbb{R}$. Nevertheless, in the remainder of this work, a pessimistic scenario with equiprobable input data are assumed. Therefore, $\lambda^1_{AVG}$ is obtained from Equations (\ref{gather:avg_latency}) and (\ref{gather:phit_equiprob}), and the resulting values are reported in Table \ref{tab:average_latency}. In comparison, a conventional pipelined quire adder with $w$-bit fragments ~\cite{sharma2023CLARINET} will always need $L + n/w-1$ cycles. So even in the worst cases where there are many fragments, i.e. the first row of Table~\ref{tab:average_latency}, there is a noticeable gain. Consider for instance $n=1024, w=2$: the penalty in the conventional pipelined case would be $n/w-1 = 511$ cycles, whereas in the speculative case such penalty would be $\lambda_{AVG}^{1} - 1 = 8.6$ cycles.

\subsubsection{Integrating the \ac{MSADD} into the \ac{PositMAC}}
\label{subsubsec:Integrating the Multispeculative Adder into the Posit MAC}

    \begin{figure*}[!b]
        \centering
        \includegraphics[width=1.0\linewidth]{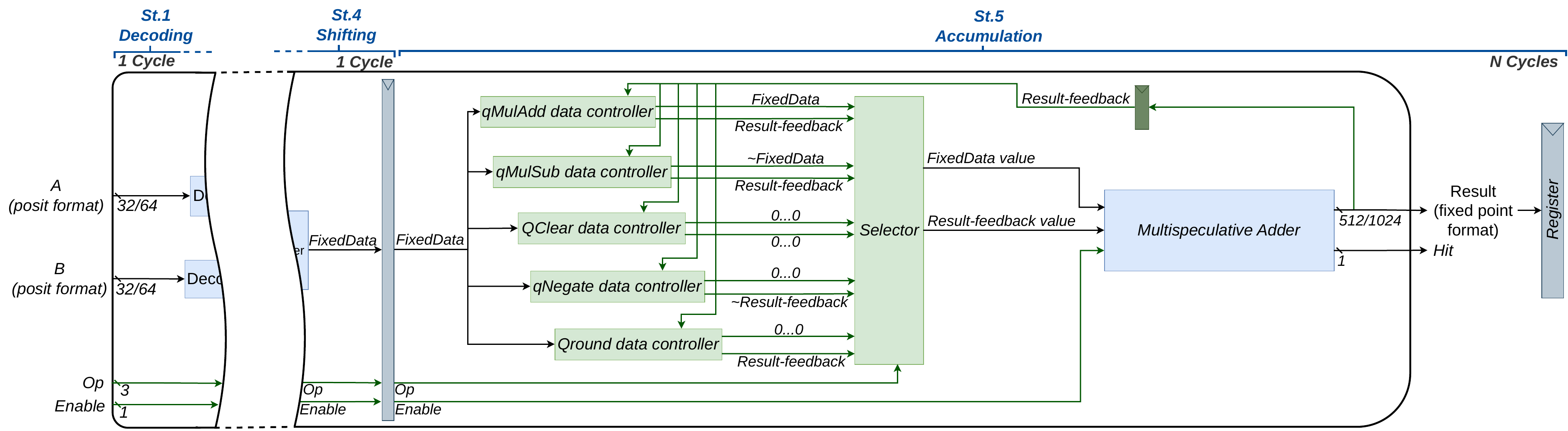}
        
        \vspace{-0.5em}
        \label{fig:positmac_ms_stage5}
        \caption{Implementation of the \ac{MSADD} in Stage~5 of the Multispeculative \ac{PositMAC} architecture. The elements highlighted in green correspond to the additional control circuitry.}
    \end{figure*}

    \noindent
    The proposed \ac{PositMAC} incorporates the simplified \ac{MSADD} shown in Fig. \ref{fig:msadd_new} into Stage~5 of the pipeline, replacing the baseline monolithic quire adder, as illustrated in Fig. \ref{fig:positmac_ms_stage5}. Due to the structure of the \ac{MSADD}, the \emph{speculation phase} may require more than one clock cycle to complete. Consequently, the restructured Posit MAC pipeline must be modified to handle this situation. Furthermore, this requires the introduction of an additional operation, denoted \emph{\texttt{QRound}}, to notify the unit that the sequence of multiply-accumulate operations has completed, allowing the adder to transition from the \emph{accumulation phase} to the \emph{speculative phase}.

    \begin{table}[tb]
      \centering
      \caption{Average speculation latency $\lambda^{1}_{AVG}$ as a function of
      the number of fragments $n/w$. The value is independent of the posit size;
      the last two columns list the fragment width $w$ that yields each case for
      the PositMAC32 ($n=512$) and PositMAC64 ($n=1024$) units.}
       \label{tab:average_latency}
      \begin{tabular}{ccccc}
        \toprule
        $n/w$ & $\lambda^{1}_{AVG}$ (cycles) & $w$ (PositMAC32) & $w$ (PositMAC64) \\
        \midrule
        512 & 9.60 & --  & 2   \\
        256 & 8.59 & 2   & 4   \\
        128 & 7.56 & 4   & 8   \\
        64  & 6.52 & 8   & 16  \\
        32  & 5.45 & 16  & 32  \\
        16  & 4.37 & 32  & 64  \\
        8   & 3.30 & 64  & 128 \\
        4   & 2.30 & 128 & 256 \\
        2   & 1.50 & 256 & 512 \\
        \bottomrule
      \end{tabular}
    \end{table}
    
    The multicycle behaviour introduced by the speculation phase makes it necessary to characterise the latency of the complete Posit MAC unit before evaluating it experimentally. Furthermore, given that the proposed design is pipelined, the depth of the pipeline ($D$) must also be considered. The restructured PositMAC comprises five stages, four of which precede the MSADD located in Stage~5. Therefore, the average latency of a kernel containing $L$ MAC operations is depicted by Equation~\ref{gather:avg_latency_sequence_piped}.

    \begin{equation}
        \label{gather:avg_latency_sequence_piped}
        \begin{aligned}
        Latency_{MAC}(L) &= D - 1 + \lambda_{\text{AVG}}^{L} ~, \\
        &= D - 1 + L + \lambda_{\text{AVG}}^{1} - 1 ~, \\
        &= 5 - 1 + L + \lambda^{1}_{AVG}-1 ~, \\
        &= L + 3 + \lambda^{1}_{AVG}.
        \end{aligned}
    \end{equation}

    where the term $\lambda^{1}_{AVG}$ accounts for the additional speculation cycles required for the final accumulation, once the \texttt{QRound} operation has triggered the transition from the accumulation phase to the speculative phase. 
    
    The average speculation penalty $\lambda^{1}_{AVG}$ depends only on the number of internal fragments $n/w$. Therefore, \ac{PositMAC32} and \ac{PositMAC64} configurations having the same $n/w$ exhibit the same average speculation latency, although this condition corresponds to different fragment widths $w$, as shown in Table~\ref{tab:average_latency}. 
    
    Simply replacing the original quire adder with a \ac{MSADD} is possible; however, this approach introduces several issues related to both operation management and pipeline execution. For example, the input parameter \emph{C} cannot be synchronized with the pipeline stages, as it is injected directly into Stage~5 instead of propagating through the pipeline from Stage~1. Furthermore, the \emph{\texttt{qNegate}} operation requires more clock cycles than necessary because the control of the input parameters is handled exclusively by the top-level unit of the \acs{PAU}~\cite{mallasen2022PERCIVAL}.
    
    To overcome these limitations, the internal design of Stage~5 has been modified by incorporating additional control logic to manage both the input parameters of the \ac{MSADD} and the feedback of the current quire value. The latter is enabled by the registers newly introduced within the \ac{MSADD} architecture. As a result, part of the control circuitry originally implemented in the top-level unit has been relocated into Stage~5.
    
    These modifications also affect the unit interface. In particular, the input parameter \emph{C} has been removed and replaced by an internal feedback path, while several new control input signals have been introduced to specify the operation to be performed. Although these changes increase the area and power consumption of the Posit MAC unit itself, they are expected to reduce the hardware overhead of the top-level unit by simplifying its control logic. Fig.~\ref{fig:positmac_ms_stage5} illustrates the integration of the \ac{MSADD} into Stage~5 of the \ac{PositMAC} architecture.

\section{Experiments}
\label{sec:experiments}

This section presents the experimental evaluation of the proposed \ac{PositMAC32} and \ac{PositMAC64} units, detailing synthesis results, performance analysis, and comparisons against state-of-the-art designs.

\subsection{Restructured pipeline synthesis}
\label{subsec:Restructured pipeline synthesis results}

Table~\ref{table:PositMAC_restructured_InitialValues_0.5ns} presents the synthesis results obtained for both the baseline architecture and the restructured pipeline versions of the \ac{PositMAC32} and \ac{PositMAC64} units, without preserving the internal component hierarchy.

Both restructured implementations exhibit increased total area due to the additional pipeline registers introduced during restructuring. However, the overhead differs significantly, increasing by \emph{12.8\%} in the \ac{PositMAC32} unit and only \emph{1.7\%} in the \ac{PositMAC64} unit. This difference stems from their original pipeline organization: the baseline \ac{PositMAC32} architecture had three stages, whereas the \ac{PositMAC64} version already employed four, with proportionally larger registers. Consequently, adding the new pipeline stages has a more pronounced relative impact on the \ac{PositMAC32} implementation.

On the other hand, total power consumption increases by \emph{38.8\%} and \emph{15.0\%} for \ac{PositMAC32} and \ac{PositMAC64}, driven by area expansion and critical-stage delay reductions of \emph{37.2\%} and \emph{35.7\%}. Nevertheless, this delay reduction yields overall energy savings of \emph{14.0\%} and \emph{26.1\%} per execution cycle.

In the restructured pipeline, Stage~2 contains the critical path in both implementations, indicating that the \emph{Multiplier} has become the performance bottleneck of the architecture and should therefore be redesigned or replaced with a faster implementation to further reduce the critical-path delay.

\begin{table}[t]
    \centering
    \caption{Synthesis results for the baseline and restructured \ac{PositMAC} pipeline, obtained using Synopsys Design Compiler with a TSMC 28\,nm standard-cell library, a \delay{0.5} timing constraint, and a flattened internal component hierarchy.}
    \label{table:PositMAC_restructured_InitialValues_0.5ns}

    \fontsize{8.5pt}{9pt}\selectfont
    \setlength{\tabcolsep}{2pt}
    \renewcommand{\arraystretch}{1.2}
    
    \def\units{false}
    \def\headerUnits{true}

    \begin{tabular}{c|cccccccc}
        \toprule
        & \textbf{Area}                 & \textbf{Power}                        & \textbf{Energy}                       & \multicolumn{5}{c}{\textbf{Delay}} \\
        & (\area{}[\headerUnits])   & (\power{}[\headerUnits])   & (\energy{}[\headerUnits])  & \multicolumn{5}{c}{(\delay{}[\headerUnits])} \\
        \cmidrule(lr){5-9}
        & & & & \textbf{St.1} & \textbf{St.2} & \textbf{St.3} & \textbf{St.4} & \textbf{St.5} \\
        \midrule

        \multirow{2}{*}{\shortstack{\textbf{\textit{Baseline}}\\\textbf{\textit{PositMAC 32}}}}
        & \multirow{2}{*}{\area{20526.66}[\units]}
        & \multirow{2}{*}{\power{23.22}[\units]}
        & \multirow{2}{*}{\energy{26.24}[\units]}
        & \multirow{2}{*}{\delay{1.13}[\units]}
        & \multirow{2}{*}{\delay{0.39}[\units]}
        & \multirow{2}{*}{\delay{0.52}[\units]}
        & \multirow{2}{*}{\delay{--}[\units]}
        & \multirow{2}{*}{\delay{--}[\units]} \\
        & & & & & & & & \\

        \multirow{2}{*}{\shortstack{\textbf{\textit{Restructured}}\\\textbf{\textit{PositMAC 32}}}}
        & \multirow{2}{*}{\area{23160.31}[\units]}
        & \multirow{2}{*}{\power{31.77}[\units]}
        & \multirow{2}{*}{\energy{22.55}[\units]}
        & \multirow{2}{*}{\delay{0.56}[\units]}
        & \multirow{2}{*}{\delay{0.71}[\units]}
        & \multirow{2}{*}{\delay{0.51}[\units]}
        & \multirow{2}{*}{\delay{0.41}[\units]}
        & \multirow{2}{*}{\delay{0.50}[\units]} \\
        & & & & & & & & \\

        \midrule
        
        \multirow{2}{*}{\shortstack{\textbf{\textit{Baseline}}\\\textbf{\textit{PositMAC 64}}}}
        & \multirow{2}{*}{\area{60339.13}[\units]}
        & \multirow{2}{*}{\power{59.58}[\units]}
        & \multirow{2}{*}{\energy{83.42}[\units]}
        & \multirow{2}{*}{\delay{1.40}[\units]}
        & \multirow{2}{*}{\delay{0.34}[\units]}
        & \multirow{2}{*}{\delay{0.32}[\units]}
        & \multirow{2}{*}{\delay{0.57}[\units]}
        & \multirow{2}{*}{\delay{--}[\units]} \\
        & & & & & & & & \\

        \multirow{2}{*}{\shortstack{\textbf{\textit{Restructured}}\\\textbf{\textit{PositMAC 64}}}}
        & \multirow{2}{*}{\area{61353.05}[\units]}
        & \multirow{2}{*}{\power{68.52}[\units]}
        & \multirow{2}{*}{\energy{61.67}[\units]}
        & \multirow{2}{*}{\delay{0.73}[\units]}
        & \multirow{2}{*}{\delay{0.90}[\units]}
        & \multirow{2}{*}{\delay{0.49}[\units]}
        & \multirow{2}{*}{\delay{0.46}[\units]}
        & \multirow{2}{*}{\delay{0.47}[\units]} \\
        & & & & & & & & \\

        \bottomrule
    \end{tabular}
\end{table}

\subsection{Optimized multiplier synthesis}
\label{subsec:Optimized multiplier synthesis results}

\begin{table}[!t]
    \centering
    \caption{Synthesis results for different multiplier implementations within the restructured PositMAC architecture. Data reported usingSynopsys Design Compiler with TSMC 28\,nm standard-cell library, \delay{0.5} timing constraint and a flattened internal component hierarchy.}
    \label{table:PositMAC_Ex2_Multiplier_Optimization}

    \fontsize{8.5pt}{9pt}\selectfont
    \setlength{\tabcolsep}{2pt}
    \renewcommand{\arraystretch}{1.2}

    \def\units{false}
    \def\headerUnits{true}

    \subfloat[Restructured PositMAC 32 synthesis results.]{%

        \begin{tabular}{l|cccccccc}
            \toprule
            & \textbf{Area}                 & \textbf{Power}                        & \textbf{Energy}                       & \multicolumn{5}{c}{\textbf{Delay}} \\
            & (\area{}[\headerUnits])   & (\power{}[\headerUnits])   & (\energy{}[\headerUnits])  & \multicolumn{5}{c}{(\delay{}[\headerUnits])} \\
            \cmidrule(lr){5-9}
            & & & & \textbf{St.1} & \textbf{St.2} & \textbf{St.3} & \textbf{St.4} & \textbf{St.5} \\
            \midrule

            \textbf{\textit{Baseline}}
            & \area{23160.31}[\units]
            & \power{31.77}[\units]
            & \energy{22.55}[\units]
            & \delay{0.56}[\units]
            & \delay{0.71}[\units]
            & \delay{0.51}[\units]
            & \delay{0.41}[\units]
            & \delay{0.50}[\units] \\

            \textbf{\textit{VHDL library}}
            & \area{23213.99}[\units]
            & \power{27.51}[\units]
            & \energy{25.31}[\units]
            & \delay{0.57}[\units]
            & \delay{0.92}[\units]
            & \delay{0.52}[\units]
            & \delay{0.41}[\units]
            & \delay{0.50}[\units] \\

            \textbf{\textit{Booth-4/KS}}
            & \area{20452.82}[\units]
            & \power{39.53}[\units]
            & \energy{19.76}[\units]
            & \delay{0.50}[\units]
            & \delay{0.50}[\units]
            & \delay{0.48}[\units]
            & \delay{0.41}[\units]
            & \delay{0.50}[\units] \\

            \textbf{\textit{Booth-8/KS}}
            & \area{20284.99}[\units]
            & \power{39.33}[\units]
            & \energy{19.66}[\units]
            & \delay{0.50}[\units]
            & \delay{0.50}[\units]
            & \delay{0.50}[\units]
            & \delay{0.41}[\units]
            & \delay{0.50}[\units] \\

            \bottomrule
        \end{tabular}
    }
    \vspace{1em}

    \subfloat[Restructured PositMAC 64 synthesis results.]{%

        \begin{tabular}{l|cccccccc}
            \toprule
            & \textbf{Area}                 & \textbf{Power}                        & \textbf{Energy}                       & \multicolumn{5}{c}{\textbf{Delay}} \\            & (\area{}[\headerUnits]) & (\power{}[\headerUnits]) & (\energy{}[\headerUnits]) & \multicolumn{5}{c}{(\delay{}[\headerUnits])} \\
            \cmidrule(lr){5-9}
            & & & & \textbf{St.1} & \textbf{St.2} & \textbf{St.3} & \textbf{St.4} & \textbf{St.5} \\
            \midrule

            \textbf{\textit{Baseline}}
            & \area{61353.05}[\units]
            & \power{68.52}[\units]
            & \energy{61.67}[\units]
            & \delay{0.73}[\units]
            & \delay{0.90}[\units]
            & \delay{0.49}[\units]
            & \delay{0.46}[\units]
            & \delay{0.47}[\units] \\

            \textbf{\textit{VHDL library}}
            & \area{60853.46}[\units]
            & \power{44.28}[\units]
            & \energy{87.24}[\units]
            & \delay{0.73}[\units]
            & \delay{1.97}[\units]
            & \delay{0.49}[\units]
            & \delay{0.46}[\units]
            & \delay{0.47}[\units] \\

            \textbf{\textit{Booth-4/KS}}
            & \area{57953.95}[\units]
            & \power{105.67}[\units]
            & \energy{52.84}[\units]
            & \delay{0.50}[\units]
            & \delay{0.50}[\units]
            & \delay{0.49}[\units]
            & \delay{0.46}[\units]
            & \delay{0.47}[\units] \\

            \textbf{\textit{Booth-8/KS}}
            & \area{52853.72}[\units]
            & \power{89.18}[\units]
            & \energy{50.83}[\units]
            & \delay{0.55}[\units]
            & \delay{0.57}[\units]
            & \delay{0.49}[\units]
            & \delay{0.46}[\units]
            & \delay{0.47}[\units] \\

            \bottomrule
        \end{tabular}
    }
    
\end{table}

To reduce the critical-path delay, the original \emph{Multiplier} was replaced with faster implementations. Table~\ref{table:PositMAC_Ex2_Multiplier_Optimization} summarizes the synthesis results for three variants: the default VHDL  multiplication operator, \emph{Booth-4} and \emph{Booth-8} algorithms using \emph{Kogge--Stone} adders in the final addition stage.

The default VHDL multiplication operator increases Stage~2 delay by $29.6\%$ and $118.9\%$ for the \ac{PositMAC32} and \ac{PositMAC64} units, respectively, making it unsuitable. In contrast, the \emph{Booth-8/KS} implementation reduces Stage~2 delay by $29.6\%$ and $36.7\%$ for the \ac{PositMAC32} and \ac{PositMAC64} units, respectively. However, while the \ac{PositMAC32} meets the \delay{0.5} timing constraint, the \ac{PositMAC64} still exceeds it, with a minimum delay of \delay{0.57}.

Finally, the \emph{Booth-4/KS} implementation provides a larger Stage~2 delay reduction of $44.4\%$ for the \emph{PositMAC64} unit while satisfying the \delay{0.5} timing constraint in both units, making it the preferred choice for subsequent experiments. It reduces area by $11.7\%$ and $5.5\%$ for \ac{PositMAC32} and \emph{PositMAC64}, respectively, but increases total power consumption by $24.4\%$ and $54.2\%$. Offsetting this, the shorter critical delay lowers single-cycle energy consumption by $12.4\%$ and $14.3\%$. Nevertheless, despite this substantial delay reduction, the \emph{Multiplier} component remains the architecture's critical path.

\subsection{Multispeculative PositMAC synthesis}
\label{subsec: Multispeculative Adder implementation synthesis results}

\begin{table*}[!b]
    \centering

    \caption{Synthesis results comparing the restructured baseline architecture (\emph{Monolithic}), implementing the \emph{Booth-4/KS} multiplier and the original monolithic adder, with different \emph{Multispeculative PositMAC} implementations employing \ac{MSADD} units based on \emph{Kogge--Stone}, \emph{Brent--Kung}, and \emph{Ripple-Carry} adder architectures with different group sizes (\emph{w}). All designs were synthesized using Synopsys Design Compiler under a \delay{0.5} timing constraint. \emph{Red.} denotes the percentage reduction achieved by each \ac{MSADD}-based Posit \ac{MAC} implementation with respect to the corresponding monolithic Posit MAC architecture.}
    \label{table:PositMAC_MS_ImplementingMSADD_v2}

    \fontsize{8.5pt}{8pt}\selectfont
    \setlength{\tabcolsep}{5pt}
    \renewcommand{\arraystretch}{1.2}

    \def\units{false}
    \def\headerUnits{true}

    \begin{tabular}{c l !{\vrule width 1pt} c c c c !{\vrule width 1pt} c c c c !{\vrule width 1pt} c c c c}
        
        \toprule
        
        & &
        \multicolumn{4}{c}{\rule{0pt}{1.4em}\textbf{\textit{\normalsize Kogge--Stone (KS)}}}{\vrule width 1pt}&
        \multicolumn{4}{c}{\rule{0pt}{1.4em}\textbf{\textit{\normalsize Brent--Kung (BK)}}}{\vrule width 1pt}&
        \multicolumn{4}{c}{\rule{0pt}{1.4em}\textbf{\textit{\normalsize Ripple Carry (RC)}}} \\ [4pt]
        
        & &
        \shortstack{\textbf{Area}\\[-1ex]} &
        \shortstack{\textbf{(Red.)}\\[-1ex]} &
        \shortstack{\textbf{Power}\\[-1ex]} &
        \shortstack{\textbf{(Red.)}\\[-1ex]} &
        \shortstack{\textbf{Area}\\[-1ex]} &
        \shortstack{\textbf{(Red.)}\\[-1ex]} &
        \shortstack{\textbf{Power}\\[-1ex]} &
        \shortstack{\textbf{(Red.)}\\[-1ex]} &
        \shortstack{\textbf{Area}\\[-1ex]} &
        \shortstack{\textbf{(Red.)}\\[-1ex]} &
        \shortstack{\textbf{Power}\\[-1ex]} &
        \shortstack{\textbf{(Red.)}\\[-1ex]} \\
        
        \cmidrule(lr){3-4}
        \cmidrule(lr){5-6}
        \cmidrule(lr){7-8}
        \cmidrule(lr){9-10}
        \cmidrule(lr){11-12}
        \cmidrule(lr){13-14}

        & &
        \area{}[\headerUnits] & (\%) &
        \power{}[\headerUnits] & (\%) &
        \area{}[\headerUnits] & (\%) &
        \power{}[\headerUnits] & (\%) &
        \area{}[\headerUnits] & (\%) &
        \power{}[\headerUnits] & (\%) \\
        
        \midrule


        \multirow{9}{*}{\rotatebox[origin=c]{90}{\textbf{Posit MAC 32}}} &
        \textbf{\textit{MSADD w=2}} & 
        \area{15372.63}[\units] & 
        (24.8)&
        \power{29.82}[\units] & 
        (24.6)&

        \area{15372.63}[\units] & 
        (18.0)&
        \power{29.82}[\units] & 
        (23.5)&

        \area{15374.77}[\units] & 
        (19.4)&
        \power{29.77}[\units] & 
        (24.2) \\

        &
        \textbf{\textit{MSADD w=4}} & 
        \area{14945.74}[\units] & 
        (26.9)&
        \power{29.23}[\units] & 
        (26.1)&
    
        \area{14945.74}[\units] & 
        (20.3)&
        \power{29.23}[\units] & 
        (25.0)&

        \area{14969.93}[\units] & 
        (21.6)&
        \power{29.18}[\units] & 
        (25.8) \\

        &
        \textbf{\textit{MSADD w=8}} & 
        \area{14884.76}[\units] & 
        (27.2)&
        \power{29.09}[\units] & 
        (26.4)&

        \area{14884.76}[\units] & 
        (20.6)&
        \power{29.09}[\units] & 
        (25.3)&

        \area{14947.88}[\units] & 
        (21.7)&
        \power{29.33}[\units] & 
        (25.4) \\

        &
        \textbf{\textit{MSADD w=16}} & 
        \area{15330.67}[\units] & 
        (25.0)&
        \power{29.26}[\units] & 
        (26.0)&

        \area{14934.78}[\units] & 
        (20.4)&
        \power{28.84}[\units] & 
        (26.0)&

        \area{16551.36}[\units] & 
        (13.3)&
        \power{31.23}[\units] & 
        (20.5) \\

        &
        \textbf{\textit{MSADD w=32}} & 
        \area{15776.08}[\units] & 
        (22.9)&
        \power{29.60}[\units] & 
        (25.1)&

        \area{15137.39}[\units] & 
        (19.3)&
        \power{28.86}[\units] & 
        (25.9)&

        \area{16249.59}[\units] & 
        (14.9)&
        \power{30.32}[\units] & 
        (22.9) \\

        &
        \textbf{\textit{MSADD w=64}} & 
        \area{16264.46}[\units] & 
        (20.5)&
        \power{29.48}[\units] & 
        (25.4)&

        \area{15453.65}[\units] & 
        (17.6)&
        \power{29.18}[\units] & 
        (25.1)&
        
        \area{16336.91}[\units] & 
        (14.4)&
        \power{29.65}[\units] & 
        (24.6) \\

        &
        \textbf{\textit{MSADD w=128}} & 
        \area{16650.77}[\units] & 
        (18.6)&
        \power{29.63}[\units] & 
        (25.0)&

        \area{15946.06}[\units] & 
        (15.0)&
        \power{28.82}[\units] & 
        (26.0)&

        \area{16478.28}[\units] & 
        (13.7)&
        \power{29.40}[\units] & 
        (25.2) \\

        &
        \textbf{\textit{MSADD w=256}} & 
        \area{17396.69}[\units] & 
        (14.9)&
        \power{29.92}[\units] & 
        (24.3)&

        \area{16216.33}[\units] & 
        (13.5)&
        \power{29.22}[\units] & 
        (25.0)&

        \area{16232.33}[\units] & 
        (15.0)&
        \power{29.10}[\units] & 
        (29.1) \\

        &
        \textbf{\textit{Monolithic}} & 
        \area{20452.82}[\units] & 
        (0.0)&
        \power{39.53}[\units] & 
        (0.0)&

        \area{18753.59}[\units] & 
        (0.0)&
        \power{38.96}[\units] & 
        (0.0)&

        \area{19085.22}[\units] & 
        (0.0)&
        \power{39.30}[\units] & 
        (0.0) \\
        

        \midrule


        \multirow{10}{*}{\rotatebox[origin=c]{90}{\textbf{Posit MAC 64}}} &
        \textbf{\textit{MSADD w=2}} & 
        \area{45305.57}[\units] & 
        (21.8)&
        \power{83.60}[\units] & 
        (20.9)&

        \area{45305.57}[\units] & 
        (14.6)&
        \power{83.60}[\units] & 
        (19.4)&

        \area{49095.14}[\units] & 
        (8.5)&
        \power{92.74}[\units] & 
        (11.7) \\

        &
        \textbf{\textit{MSADD w=4}} & 
        \area{44790.35}[\units] & 
        (22.7)&
        \power{82.84}[\units] & 
        (21.6)&

        \area{44790.35}[\units] & 
        (15.5)&
        \power{82.84}[\units] & 
        (20.1)&

        \area{48996.49}[\units] & 
        (8.7)&
        \power{89.62}[\units] & 
        (14.7) \\

        &
        \textbf{\textit{MSADD w=8}} & 
        \area{45375.37}[\units] & 
        (21.7)&
        \power{83.42}[\units] & 
        (21.0)&

        \area{45375.37}[\units] & 
        (14.4)&
        \power{83.42}[\units] & 
        (19.5)&

        \area{49816.12}[\units] & 
        (7.2)&
        \power{89.18}[\units] & 
        (15.1) \\

        &
        \textbf{\textit{MSADD w=16}} & 
        \area{46397.48}[\units] & 
        (19.9)&
        \power{84.55}[\units] & 
        (20.0)&

        \area{47446.43}[\units] & 
        (10.5)&
        \power{87.63}[\units] & 
        (15.5)&

        \area{50388.16}[\units] & 
        (6.1)&
        \power{89.41}[\units] & 
        (14.9) \\

        &
        \textbf{\textit{MSADD w=32}} & 
        \area{48746.75}[\units] & 
        (15.9)&
        \power{86.83}[\units] & 
        (17.8)&

        \area{48423.19}[\units] & 
        (8.7)&
        \power{87.24}[\units] & 
        (15.8)&

        \area{51252.39}[\units] & 
        (4.5)&
        \power{89.19}[\units] & 
        (15.1) \\

        &
        \textbf{\textit{MSADD w=64}} & 
        \area{49364.03}[\units] & 
        (14.8)&
        \power{85.64}[\units] & 
        (19.0)&

        \area{47722.37}[\units] & 
        (10.0)&
        \power{85.02}[\units] & 
        (18.0)&

        \area{50706.56}[\units] & 
        (5.5)&
        \power{88.98}[\units] & 
        (15.3) \\

        &
        \textbf{\textit{MSADD w=128}} & 
        \area{51291.58}[\units] & 
        (11.5)&
        \power{87.25}[\units] & 
        (17.4)&

        \area{47390.11}[\units] & 
        (10.6)&
        \power{84.17}[\units] & 
        (18.8)&

        \area{50561.53}[\units] & 
        (5.8)&
        \power{88.50}[\units] & 
        (15.8) \\

        &
        \textbf{\textit{MSADD w=256}} & 
        \area{52206.84}[\units] & 
        (9.9)&
        \power{86.07}[\units] & 
        (18.5)&

        \area{47862.86}[\units] & 
        (9.7)&
        \power{83.85}[\units] & 
        (19.1)&

        \area{49319.55}[\units] & 
        (8.1)&
        \power{87.35}[\units] & 
        (16.9) \\

        &
        \textbf{\textit{MSADD w=512}} & 
        \area{53822.41}[\units] & 
        (7.1)&
        \power{87.79}[\units] & 
        (16.9)&

        \area{48235.07}[\units] & 
        (9.0)&
        \power{84.69}[\units] & 
        (18.3)&

        \area{49696.42}[\units] & 
        (8.1)&
        \power{87.15}[\units] & 
        (17.1) \\

        &
        \textbf{\textit{Monolithic}} & 
        \area{57953.95}[\units] & 
        (0.0)&
        \power{105.67}[\units] & 
        (0.0)&      

        \area{53023.70}[\units] & 
        (0.0)&
        \power{103.66}[\units] & 
        (0.0)&

        \area{53678.65}[\units] & 
        (0.0)&
        \power{105.08}[\units] & 
        (0.0) \\


        \bottomrule
        
    \end{tabular}

\end{table*}

To reduce the area and power consumption of the Posit MAC units, the original \emph{Quire Adder} component is replaced with the proposed \emph{Multispeculative Adder}. Table~\ref{table:PositMAC_MS_ImplementingMSADD_v2} presents the synthesis results for the different \ac{MSADD}-based \ac{PositMAC} implementations and compares them with those of the corresponding restructured baseline architecture.

The average area reduction achieved by the \emph{MSADD/KS} implementations is $22.6\%$ and $16.2\%$ for the \ac{PositMAC32} and \ac{PositMAC64} units, respectively. The corresponding reductions for the \emph{MSADD/BK} implementations are $18.1\%$ and $11.5\%$, while the \emph{MSADD/RC} implementations achieve average reductions of $16.7\%$ and $11.7\%$. Among the evaluated architectures, the \emph{Kogge--Stone} implementation provides the largest percentual average area reduction. 

The maximum area reduction is achieved with an internal fragment size of $w=8$ for the \ac{PositMAC32} unit and $w=4$ for the \ac{PositMAC64} unit. A similar trend is observed for the other adder architectures, with the maximum area reduction obtained for the same or comparable internal fragment sizes. Regarding the absolute area, the lowest values are achieved by the \emph{Kogge--Stone} and \emph{Brent--Kung} implementations, with $w=8$ for the \ac{PositMAC32} unit and $w=4$ for the \ac{PositMAC64} unit. In general, the results indicate that increasing the internal fragment size reduces the area savings achieved by the \emph{Multispeculative Adder}. 

Regarding power consumption, the \emph{MSADD/KS} implementations reduce the average power consumption by $25.4\%$ and $19.3\%$ for the \ac{PositMAC32} and \ac{PositMAC64} units, respectively. The \emph{MSADD/BK} implementations achieve comparable reductions of $25.2\%$ and $18.3\%$, whereas the \emph{MSADD/RC} implementations provide average reductions of $24.3\%$ and $18.8\%$. Consistent with the area results, the \emph{Kogge--Stone} implementation provides the highest average power reduction in percentage, with the maximum reduction achieved using an internal group size of $w=8$ for the \ac{PositMAC32} unit and $w=4$ for the \ac{PositMAC64} unit. In absolute terms, for the \ac{PositMAC32} unit, the lowest absolute power consumption is obtained with the \emph{Brent--Kung} implementation and an internal group size of $w=128$, while for the \ac{PositMAC64} unit, the lowest value is achieved by both the \emph{Kogge--Stone} and \emph{Brent--Kung} implementations with $w=4$. 
    
    \subsubsection{Synthesis results for MSADD with additional control circuitry}
    \label{subsubsec: Synthesis Results for Multispeculative Adder with Additional Control Circuitry}
    
    \begin{figure*}[!t]
        \centering
        
        \includegraphics[width=\linewidth]{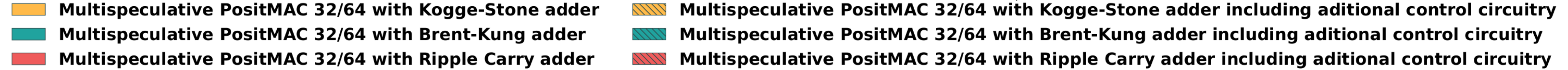}%
        
        \vspace{1em}
        
        \subfloat[Multispeculative PositMAC32 total area comparison.\label{fig:multispeculative_positMAC_comparison:PositMAC32_area}]{%
            \includegraphics[width=0.49\linewidth]{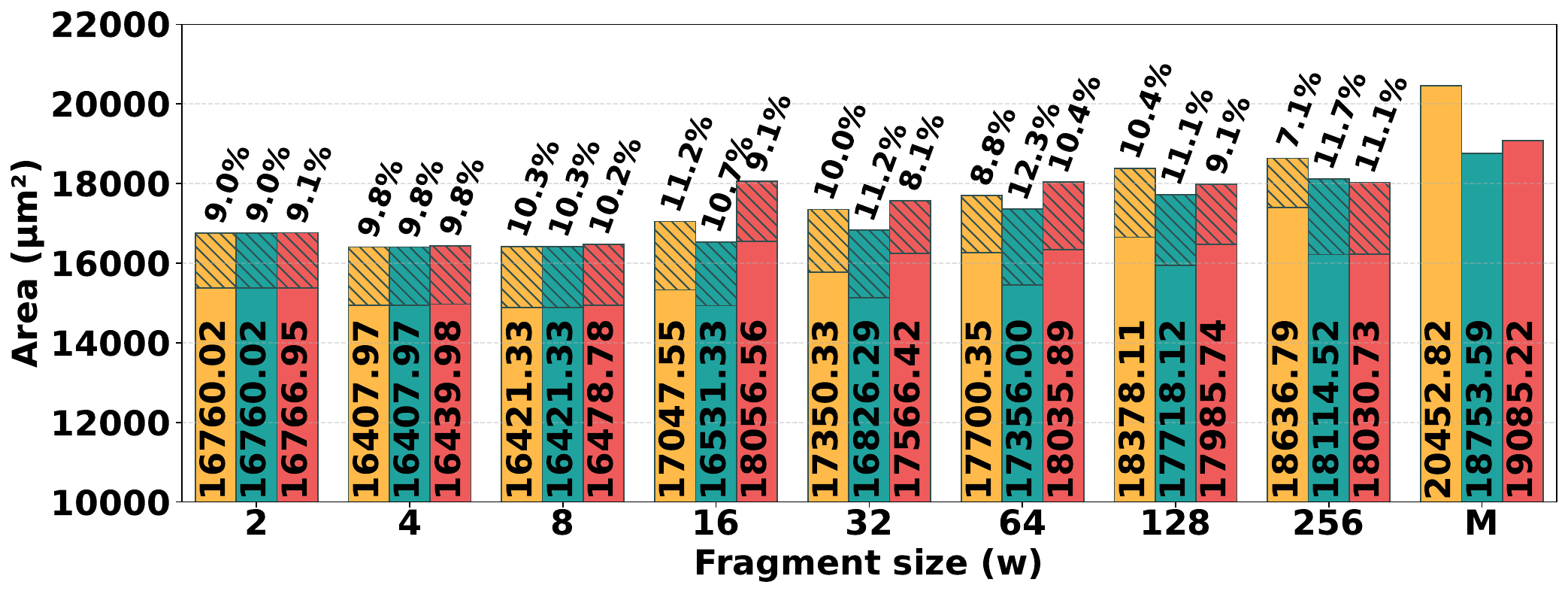}%
        }
        \hfill
        \subfloat[Multispeculative PositMAC64 total area comparison.\label{fig:multispeculative_positMAC_comparison:PositMAC64_area}]{%
            \includegraphics[width=0.49\linewidth]{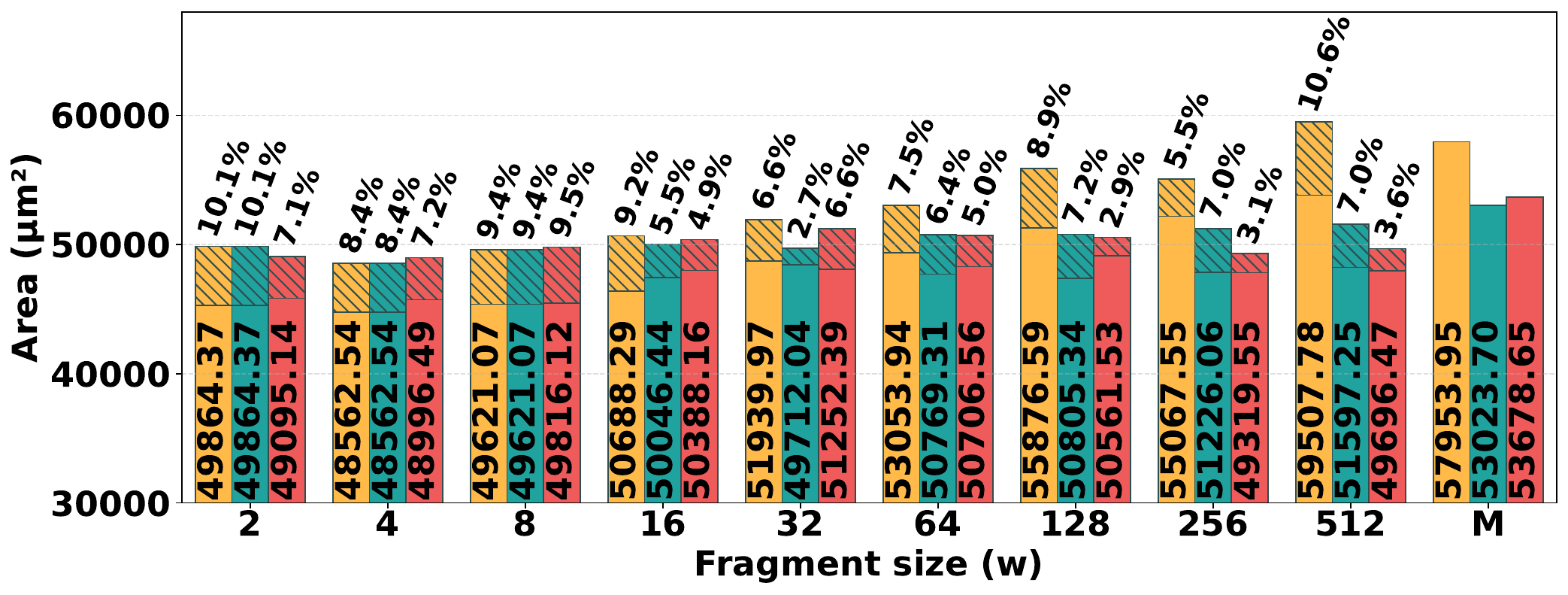}%
        }

        \subfloat[Multispeculative PositMAC32 total power comparison.\label{fig:multispeculative_positMAC_comparison:PositMAC32_power}]{%
            \includegraphics[width=0.49\linewidth]{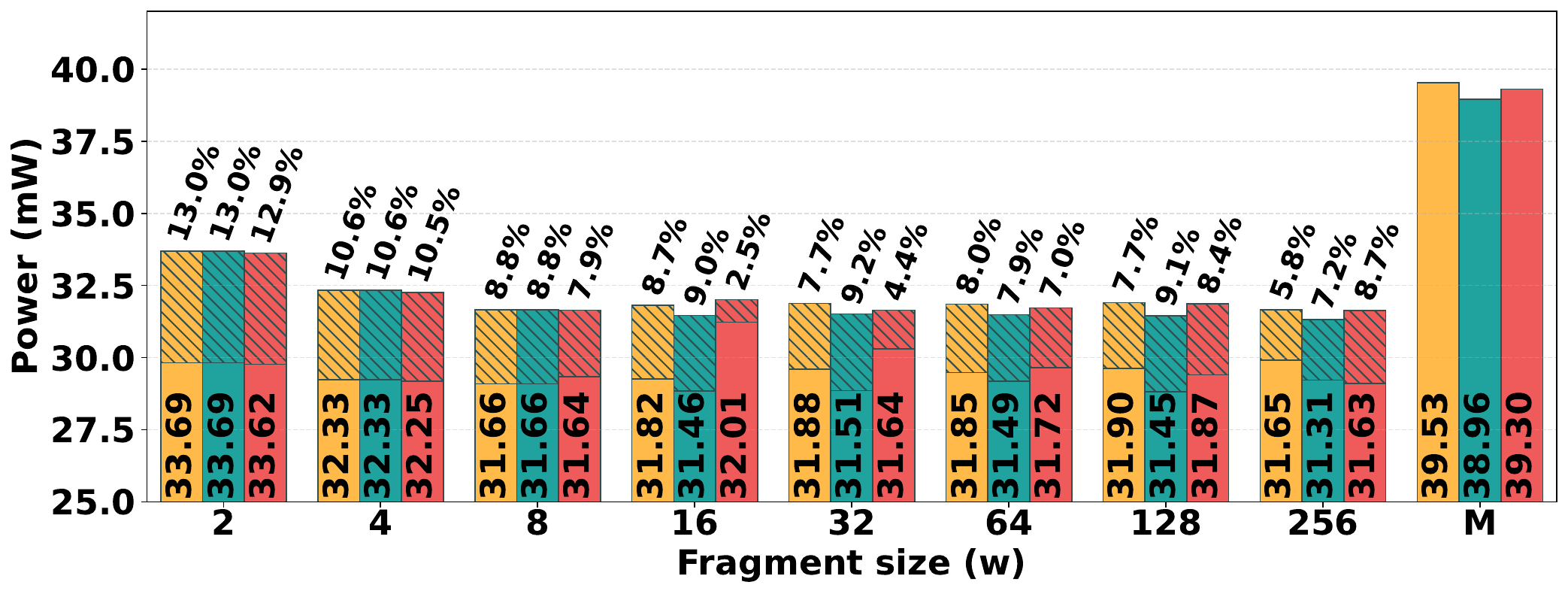}%
        }
        \hfill
        \subfloat[Multispeculative PositMAC64 total power comparison.\label{fig:multispeculative_positMAC_comparison:PositMAC64_power}]{%
            \includegraphics[width=0.49\linewidth]{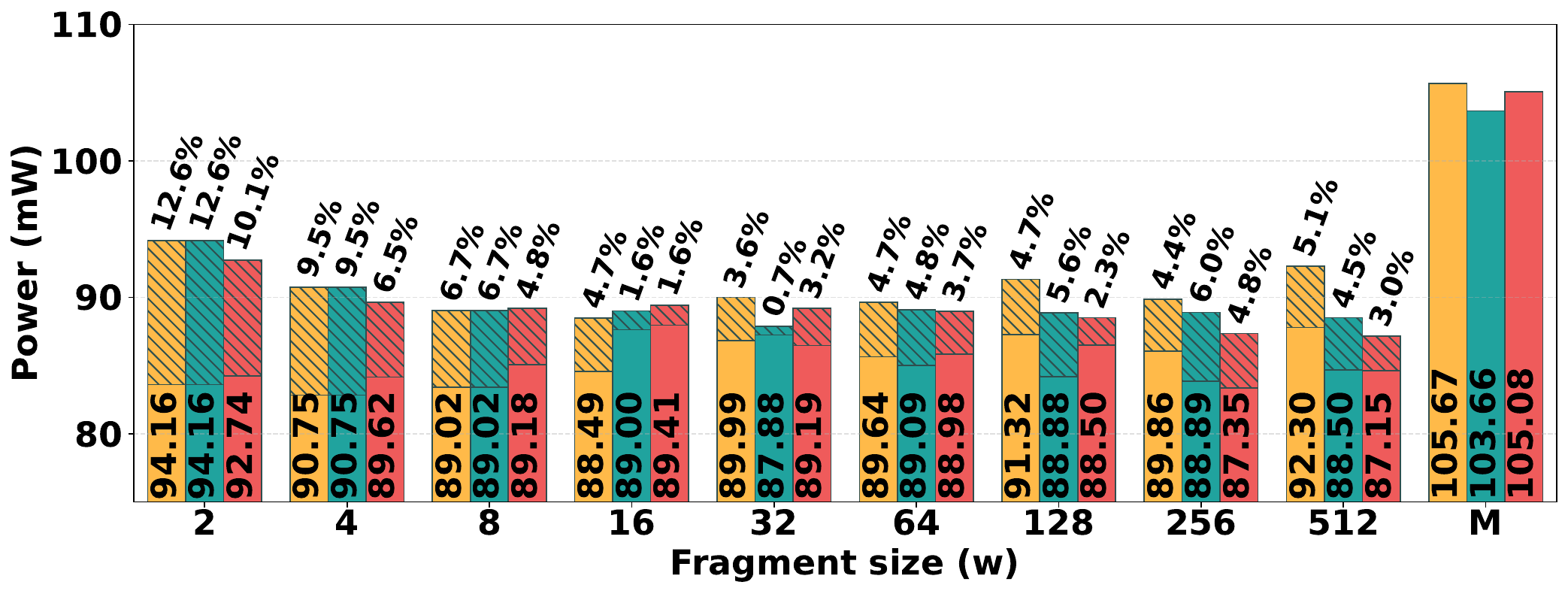}%
        }
        
        \caption{\emph{Multispeculative PositMAC} comparison in terms of area (a--b) and power consumption (c--d). The evaluation considers the restructured baseline architecture, implementing the \emph{Booth-4/KS} multiplier and the original monolithic adder (M), together with several MSADD implementations employing different internal adder architectures (\emph{Kogge--Stone}, \emph{Brent--Kung}, and \emph{Ripple Carry}) and inner group sizes (w). All designs were synthesized using Synopsys Design Compiler under a \delay{0.5} timing constraint. The values shown below each bar correspond to the implementations with the additional control circuitry, whereas the percentages above the bars indicate the relative overhead introduced by incorporating the additional control circuitry to the corresponding \emph{PositMAC} unit.}
        \label{fig:multispeculative_positMAC_comparison}
    \end{figure*}

    To overcome the limitations discussed in Section~\ref{subsubsec:Integrating the Multispeculative Adder into the Posit MAC}, additional control circuitry is integrated into the Posit MAC units. Fig.~\ref{fig:multispeculative_positMAC_comparison} compares the area, power, and delay of the different \emph{Multispeculative PositMAC} implementations, highlighting the impact of the additional circuitry.

    The control circuitry overhead remains fairly uniform across adder architectures, averaging $10.0\%$ and $7.1\%$ for \ac{PositMAC32} and \ac{PositMAC64}, respectively. Relative overhead is lowest for the \emph{Kogge--Stone} implementation with $w=256$ in \ac{PositMAC32} and the \emph{Brent--Kung} implementation with $w=32$ in \ac{PositMAC64}. Additionally, $w=4$ implementations of either \emph{Kogge--Stone} or \emph{Brent--Kung} yield the lowest absolute area, achieving nearly identical results across both units.

    Regarding power overhead, the average increase is $8.7\%$ and $5.5\%$ for \ac{PositMAC32} and \ac{PositMAC64}, respectively. Relative overhead is lowest for the \emph{Ripple Carry} implementation with $w=16$ in the 32-bit unit and the \emph{Brent--Kung} version with $w=32$ in the 64-bit design. However, the lowest absolute power is achieved by \emph{Brent--Kung} with $w=256$ in \ac{PositMAC32} and \emph{Ripple Carry} with $w=512$ in \ac{PositMAC64}.
    
    Before adding control circuitry, smaller group sizes generally yielded lower power; however, their higher sensitivity to this overhead makes larger group sizes preferable in the final implementation. Stage~2 remains the critical path for both \ac{PositMAC32} and \ac{PositMAC64}. Consequently, changes to Stage~5 do not impact overall timing, as the \emph{Multiplier} dictates critical delay.

\subsection{Execution model of the PositMAC}
\label{subsec: Execution model of the PositMAC}

Building on the latency model introduced in Section~\ref{subsubsec:Integrating the Multispeculative Adder into the Posit MAC}
(Equation~\eqref{gather:avg_latency_sequence_piped}), this section evaluates the execution latency and
energy of the proposed \ac{PositMAC} units when executing kernels of $L$
consecutive MAC operations, which correspond to the accumulation lengths
found in blocked matrix multiplication, convolutions and dot-product
computations. Following the reference length sequences reported for
Big-PERCIVAL~\cite{mallasen2024BigPERCIVAL}, the sweep is limited to $L \le 40$ MAC
operations.

\begin{figure}[!t]
    \centering
    \includegraphics[width=1.0\columnwidth]{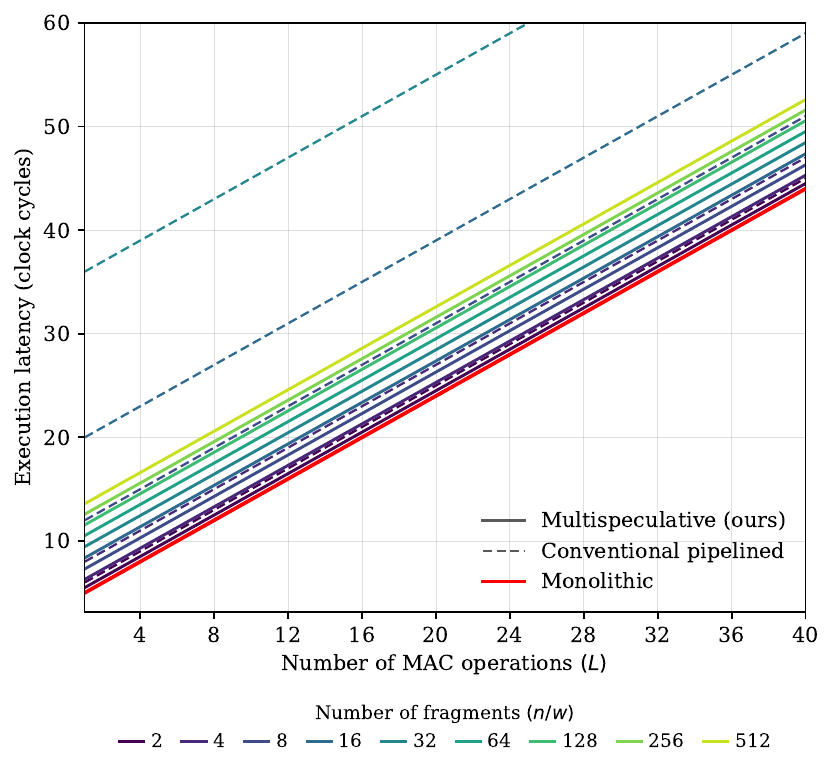}
    \vspace{-1.5em}
    
    \caption{Latency (clock cycles) of the \ac{PositMAC} units for $L \le 40$ \ac{MAC} operations, as a function of the number of fragments $n/w$. Solid lines correspond to the proposed Multispeculative designs, $Latency(L)=L+3+\lambda^{1}_{AVG}$ (Equation~\ref{gather:avg_latency_sequence_piped}); the red line is the monolithic quire adder, which does not speculate and completes in $L+4$ cycles; dashed lines of the same colour show the equivalent conventionally pipelined quire adder, whose carry propagation across the $n/w$ fragments yields $L+3+n/w$ cycles. For large numbers of fragments the conventional pipeline leaves the plotting range, evidencing the latency penalty avoided by the multispeculative accumulation.}
    
    \label{fig:positmac_total_latency}
\end{figure}

All the Multispeculative PositMAC configurations evaluated in this section were synthesized under the same $0.5$~\delay{} timing constraint. Since the selected configurations meet this target and accept one \ac{MAC} operation per cycle during the accumulation phase, a common clock period of $(T_{clk}=0.5$~\delay{} is used in the following execution model. Their sustained throughput is therefore independent of the internal adder topology and fragment width.  However, the total completion latency depends on $w$ through the final speculation term $\lambda^1_{AVG}$. As a result, the latency of all the configurations, for both the PositMAC32 and the PositMAC64 units, collapses onto nearly-parallel lines that differ only by the constant speculation penalty $\lambda^{1}_{AVG}$, shown in Table~\ref{tab:average_latency} and plotted in Fig.~\ref{fig:positmac_total_latency}. Thus, execution time is dominated by the length of the accumulation sequence, whereas the fragment width and the adder topology have a negligible impact on the sustained throughput.

\begin{figure*}[!b]
    \centering

    \begin{minipage}{0.49\linewidth}
        \centering
        \includegraphics[width=\linewidth]{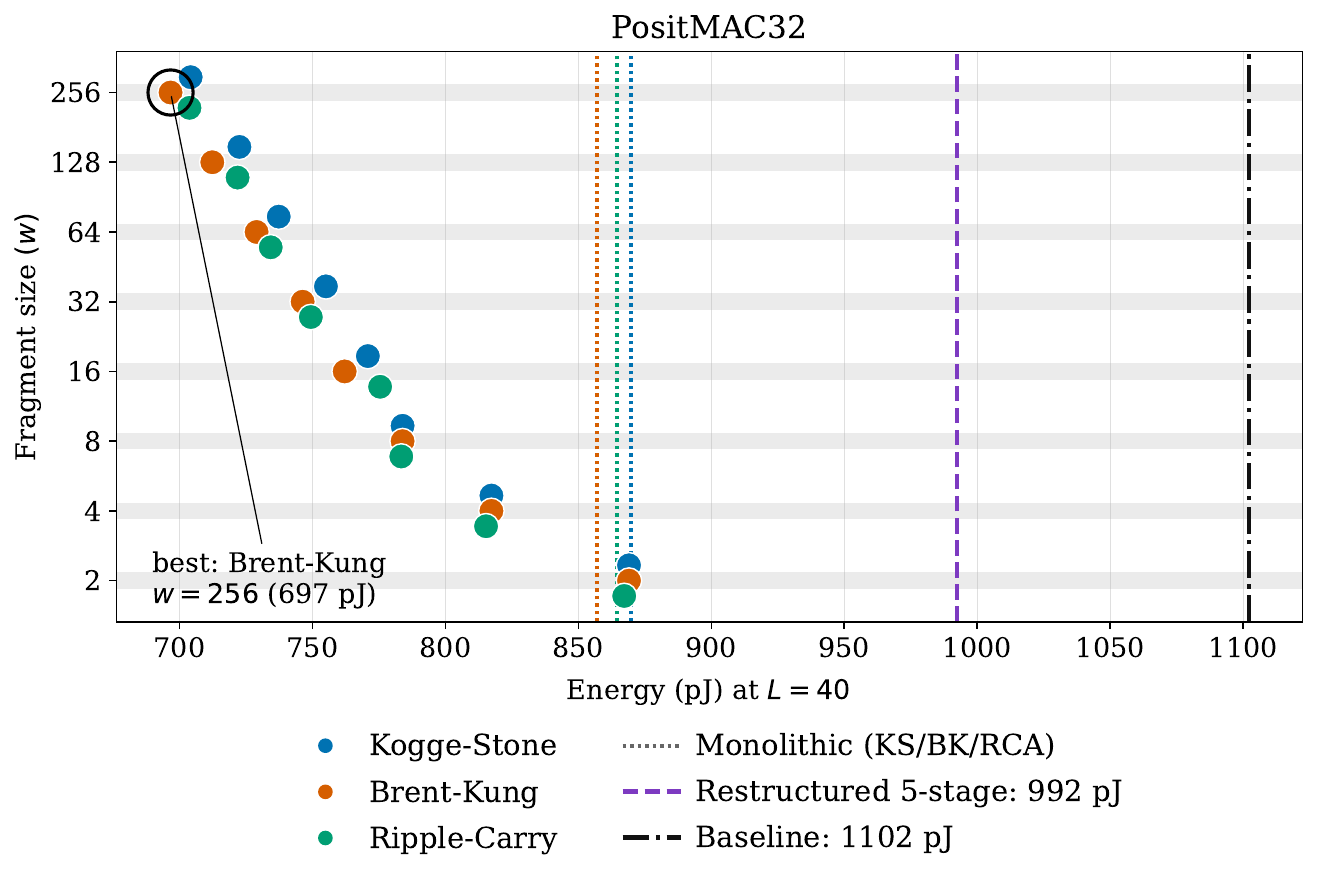}
        \label{fig:energy_PositMAC32}
    \end{minipage}
    \hfill
    \begin{minipage}{0.49\linewidth}
        \centering
        \includegraphics[width=\linewidth]{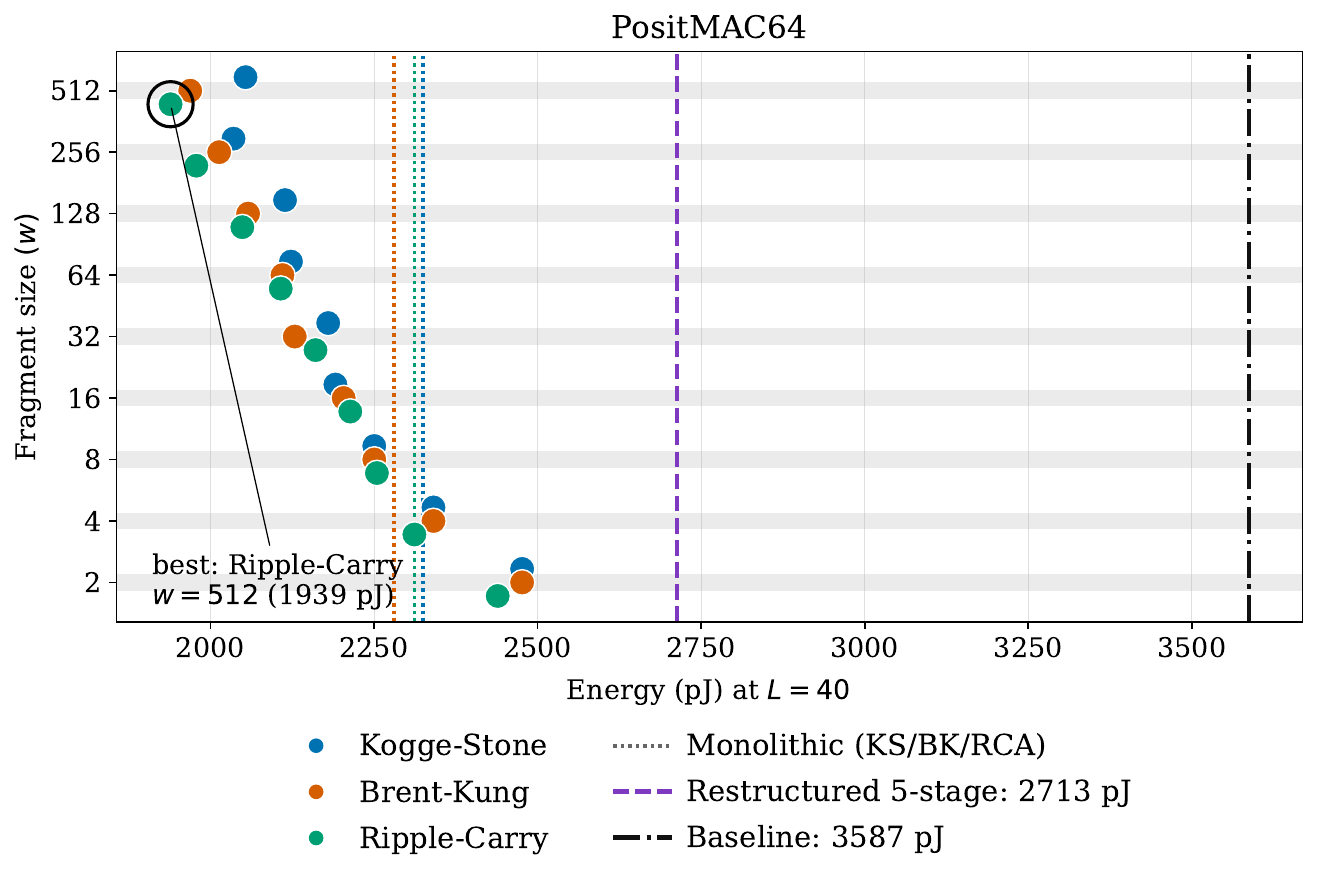}
        \label{fig:energy_PositMAC64}
    \end{minipage}
    
    \vspace{-1.5em}

    \caption{Estimated energy consumption (\energy{}) of the \ac{PositMAC} units for $L = 40$ MAC operations. Energy is computed as $E(L) = P \cdot {Latency}_{MAC}(L) \cdot T_{clk}$, where latency is computed according to Equation \ref{gather:avg_latency_sequence_piped}. $T_{clk} = 0.5$\delay{} is the target clock period used for the comparison.
    Vertical lines mark the energy of the conventional references at the same $L$: the per-quire monolithic adders (Kogge--Stone/Brent--Kung/Ripple-Carry, dotted), the restructured five-stage baseline with the default multiplier (dashed), and the original Baseline PositMAC (dash-dot). Configurations to the left of a reference are more energy-efficient than it.}
    \label{fig:execution_model_energy}
\end{figure*}

\begin{figure*}[!b]
    \centering

    \begin{minipage}{0.49\linewidth}
        \centering
        \includegraphics[width=\linewidth]{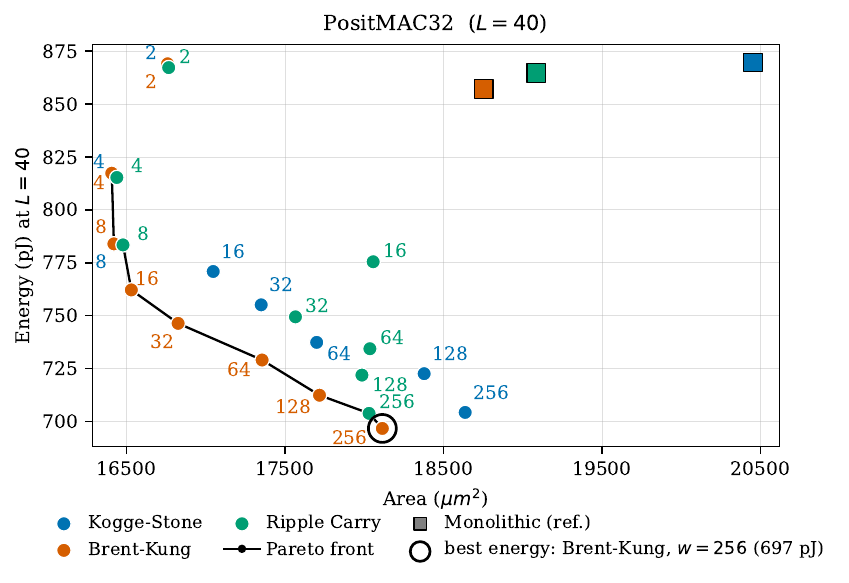}
        \label{fig:energy_area_positmac32}
    \end{minipage}
    \hfill
    \begin{minipage}{0.49\linewidth}
        \centering
        \includegraphics[width=\linewidth]{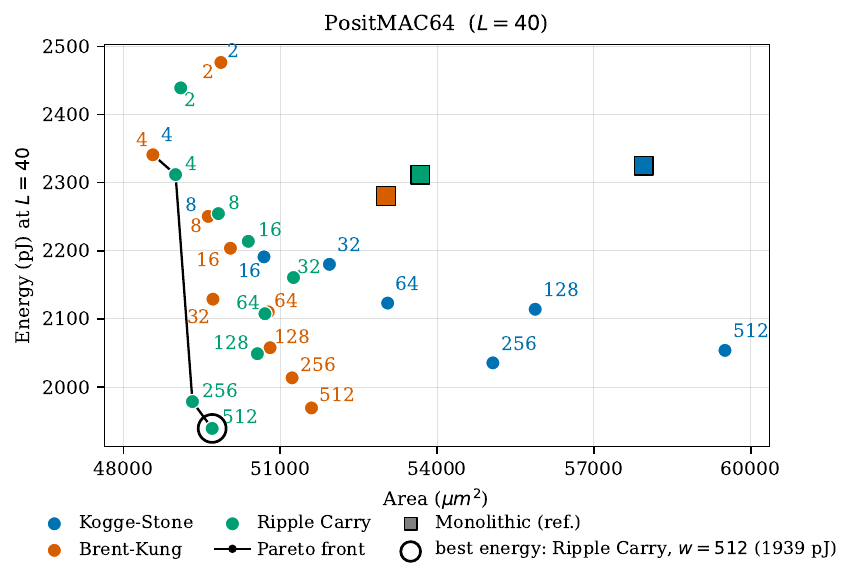}
        \label{fig:energy_area_positmac64}
    \end{minipage}
    
    \vspace{-1.5em}

    \caption{Energy--area trade-off of the evaluated implementations at a fixed accumulation length ($L=40$). Each point is a configuration (adder topology $\times$ fragment width $w$); the lower-left region is preferable and the solid line traces the Pareto front, with the best-energy configuration highlighted. The monolithic \emph{Quire Adders} (per topology) are shown as squares using the same $L+4$-cycle latency model as in Fig. \ref{fig:execution_model_energy}.}
    \label{fig:energy_area}
\end{figure*}

For comparison, the monolithic \ac{PositMAC} follows the same five-stage pipeline but does not speculate, so a sequence of $L$ MAC operations takes $L+4$ cycles (44 for $L=40$); this latency is used for the monolithic energy references in Figs.~\ref{fig:execution_model_energy} and~\ref{fig:energy_area}. Figs.~\ref{fig:positmac_total_latency} also plots, for each $n/w$, the latency of an equivalent conventionally pipelined quire adder, which must propagate the carry across its $n/w$ fragments and therefore needs $L+3+n/w$ cycles. Whereas this penalty grows with the number of fragments, rapidly leaving the plotting range, the multispeculative design stays within a few cycles of the monolithic reference, illustrating the benefit of the speculative accumulation.

In contrast to latency, energy consumption depends on the selected adder architecture and fragment width, since it combines the power with the time required to complete the MAC sequence. Fig.~\ref{fig:execution_model_energy} shows the estimated energy consumption of the PositMAC32 and PositMAC64 units for $L = 40$. Since the latency of the multispeculative designs differs only by a small constant speculation overhead, the main differences are mostly determined by their power. To put these results in context, Fig.~\ref{fig:execution_model_energy} also reports three conventional execution models: the original Baseline PositMAC, its restructured five-stage version, and the restructured design with the Booth-4/KS multiplier and a monolithic quire adder (per adder topology). The proposed multispeculative units improve on all of them; at $L=40$ the best configuration lowers the energy by about $37\%$ for PositMAC32 (from $1102$ to $697$~pJ) and $46\%$ for PositMAC64 (from $3587$ to $1939$~pJ) with respect to the Baseline. For the PositMAC32 unit, the Brent--Kung implementation with $w=256$ achieves the lowest energy, as it combines the smallest average speculation penalty with the lowest power. For the PositMAC64 unit, the Ripple Carry implementation with $w=512$ is the most energy-efficient for the same reasons. In both cases, increasing the
fragment width progressively improves the energy efficiency without affecting the sustained throughput, making the wider-fragment configurations the preferred choice for energy-constrained implementations.

\begin{table*}[!b]
    \centering
    
    \caption{Synthesis results for different state-of-the-art PositMAC architectures. Data reported using Synopsys Design Compiler with the TSMC 28\,nm standard-cell library, a flattened internal component hierarchy, and a \delay{0.5} timing constraint. \emph{QtP} refers to the \emph{Quire-to-Post} encoder implementation.}
    \label{table:State-of-the-Art_comparison}
    
    \fontsize{7.8pt}{9pt}\selectfont
    \setlength{\tabcolsep}{1.2pt}
    \renewcommand{\arraystretch}{1.2}

    \def\units{false}
    \def\headerUnits{true}

    \begin{tabular}{lcc| cccccccc !{\vrule width 1pt} cccccccc}
        
        \toprule
        & & &
        \multicolumn{8}
            {c!{\vrule width 1pt}}
            {\textbf{\textit{\normalsize PositMAC 32}}}
        &
        \multicolumn{8}{c}
            {\textbf{\textit{\normalsize PositMAC 64}}}
        \\[4pt]

        & & & & & & & & 
        & \multicolumn{2}{c!{\vrule width 1pt}}{\textbf{L=40 Ex.}}
        & & & & & & 
        & \multicolumn{2}{c}{\textbf{L=40 Ex.}} \\
        
        \cmidrule(lr){10-11}
        \cmidrule(lr){18-19}     

        & \multirow{2}{*}{\rotatebox[origin=c]{90}{\textbf{Quire}}}      
        & \multirow{2}{*}{\rotatebox[origin=c]{90}{\textbf{QtP}}}        
        & \multirow{2}{*}{\rotatebox[origin=c]{90}{\textbf{Stages}}}   
        & \textbf{Area}     & \textbf{Power}    & \textbf{Energy}   & \textbf{Delay}    & \textbf{Freq.}
        & \textbf{Energy}   & \textbf{Ex. Time}
        
        & \multirow{2}{*}{\rotatebox[origin=c]{90}{\textbf{Stages}}}    
        & \textbf{Area}     & \textbf{Power}    & \textbf{Energy}   & \textbf{Delay}    & \textbf{Freq.}
        & \textbf{Energy}   & \textbf{Ex. Time}\\

        & & & & (\area{}[\headerUnits])
        & (\power{}[\headerUnits])
        & (\energy{}[\headerUnits])
        & (\delay{}[\headerUnits])
        & (\frequency{}[\headerUnits])
        & (\energy{}[\headerUnits])
        & (\delay{}[\headerUnits])
        &
        & (\area{}[\headerUnits])
        & (\power{}[\headerUnits])
        & (\energy{}[\headerUnits])
        & (\delay{}[\headerUnits])
        & (\frequency{}[\headerUnits])
        & (\energy{}[\headerUnits])
        & (\delay{}[\headerUnits])
        \\
        \midrule

        \textbf{\textit{Crespo FMA\textsuperscript{2}}~\cite{crespo2023trading}}
        & $\times$
        & $\checkmark$
        & 5
        & \area{10100.92}[\units]
        & \power{10.14}[\units]
        & \energy{9.37}[\units]
        & \delay{0.92}[\units]
        & \frequency{1.08}[\units]
        & \energy{412.17}[\units]
        & \delay{40.66}[\units]
        
        & 5
        & \area{31862.00}[\units]
        & \power{22.24}[\units]
        & \energy{44.33}[\units]
        & \delay{1.99}[\units]
        & \frequency{0.50}[\units]
        & \energy{1950.36}[\units]
        & \delay{87.69}[\units]
        \\

        \textbf{\textit{Crespo MAC\textsuperscript{2}}~\cite{crespo2023trading}}
        & $\checkmark$
        & $\checkmark$
        & 5
        & \area{25432.09}[\units]
        & \power{18.31}[\units]
        & \energy{16.45}[\units]
        & \delay{0.90}[\units]
        & \frequency{1.11}[\units]
        & \energy{723.58}[\units]
        & \delay{39.51}[\units]
        
        & 5
        & \area{63569.02}[\units]
        & \power{34.76}[\units]
        & \energy{70.87}[\units]
        & \delay{2.04}[\units]
        & \frequency{0.49}[\units]
        & \energy{3118.44}[\units]
        & \delay{89.72}[\units]
        \\

        \textbf{\textit{Deep Positron\textsuperscript{3}}~\cite{Carmichael2019a}}
        & $\checkmark$
        & $\checkmark$
        & 3
        & \area{35121.87}[\units]
        & \power{16.90}[\units]
        & \energy{20.39}[\units]
        & \delay{1.21}[\units]
        & \frequency{0.82}[\units]
        & \energy{856.53}[\units]
        & \delay{50.69}[\units]
        
        & 3
        & \area{128716.31}[\units]
        & \power{44.15}[\units]
        & \energy{105.31}[\units]
        & \delay{2.38}[\units]
        & \frequency{0.42}[\units]
        & \energy{4413.23}[\units]
        & \delay{99.96}[\units]
        \\

        \textbf{\textit{Baseline PositMAC\textsuperscript{4}}~\cite{murillo2021EnergyEfficient}}
        & $\checkmark$
        & $\times$
        & 3
        & \area{20526.66}[\units]
        & \power{23.22}[\units]
        & \energy{26.24}[\units]
        & \delay{1.13}[\units]
        & \frequency{0.88}[\units]
        & \energy{1102.16}[\units]
        & \delay{47.46}[\units]
        
        & 4
        & \area{60339.13}[\units]
        & \power{59.58}[\units]
        & \energy{83.42}[\units]
        & \delay{1.40}[\units]
        & \frequency{0.71}[\units]
        & \energy{3586.96}[\units]
        & \delay{60.20}[\units]
        \\

        \textbf{\textit{Multispeculative PositMAC}}
        & $\checkmark$
        & $\times$
        & 5
        & \area{16531.33}[\units]
        & \power{31.46}[\units]
        & \energy{15.73}[\units]
        & \delay{0.50}[\units]
        & \frequency{2.00}[\units]
        & \energy{762.05}[\units]
        & \delay{24.23}[\units]
        
        & 5
        & \area{49712.04}[\units]
        & \power{87.88}[\units]
        & \energy{43.94}[\units]
        & \delay{0.50}[\units]
        & \frequency{2.00}[\units]
        & \energy{2128.92}[\units]
        & \delay{24.22}[\units]
        \\

        \bottomrule

    \end{tabular}

    \begin{centering}
        \footnotesize
        \textsuperscript{2}~Design URL:~\url{https://github.com/hpc-ulisboa/Posit-FMA-Units}.\\
        \textsuperscript{3}~Design URL:~\url{https://github.com/craymichael/Low-Precision-EMACs}.
        \textsuperscript{4}~Design URL:~\url{https://github.com/artecs-group/PERCIVAL}.
    \end{centering}
    
\end{table*}

At $L=40$, the most energy-efficient configuration reduces the energy per kernel by approximately $18.72\%$ for the PositMAC32 unit (Brent--Kung with $w=256$) and by approximately $16.12\%$ for the PositMAC64 unit (Ripple Carry with $w=512$) with respect to the monolithic \emph{Quire Adder}. For very small fragment widths, the higher speculation latency can offset the reduction in power consumption. This effect is particularly evident for PositMAC64 with $w=2$, where all three MSADD implementations consume more energy than their corresponding monolithic counterparts. At $w=4$, the energy is already close to the monolithic reference, whereas wider fragments provide a clear energy reduction.

Fig.~\ref{fig:energy_area} depicts the energy--area trade-off at $L=40$ MAC operations. Leveraging the area generates a Pareto Front where there are solutions with smaller values of $w$ in comparison with the energy study showcased in Fig.~\ref{fig:execution_model_energy}. For 32 bits Brent--Kung designs fully occupy the Pareto Front, whereas for 64 bits there are Ripple Carry and Brent--Kung designs in the Front and nearby. Increasing the fragment width generally reduces the speculation overhead and therefore the energy, but it does not necessarily minimize area. In particular, the minimum-energy configurations are obtained with large fragment widths, whereas the smallest-area implementations are found at intermediate or small values of $w$. Therefore, no single fragment width simultaneously minimizes both metrics.

\subsection{Comparison with the State-of-the-Art}
\label{subsec: Comparison with the State-of-the-Art}

Table~\ref{table:State-of-the-Art_comparison} compares several state-of-the-art posit MAC units, including two variants from the same study~\cite{crespo2023trading}: \emph{Crespo FMA} (without quire support) and \emph{Crespo MAC} (with quire integration). Structural choices strongly influence pipeline depth across designs: while \emph{Deep Positron}~\cite{Carmichael2019a} and the \emph{Baseline PositMAC}~\cite{murillo2021EnergyEfficient} employ shallower 3- or 4-stage pipelines to minimize latency, deeper 5-stage pipelines are adopted by both Crespo units and our \emph{Multispeculative PositMAC} to optimize operating frequency. For our proposal, the reported results correspond to a balanced trade-off configuration: the \emph{Brent--Kung} implementation with $w=16$ for \ac{Posit32} and $w=32$ for \ac{Posit64}.

Furthermore, incorporating quire logic significantly increases area and power consumption; while \emph{Crespo FMA} omits the internal quire, all remaining designs incorporate it. Except for the \emph{Baseline PositMAC} and our \emph{Multispeculative PositMAC}, all evaluated architectures include the \ac{QtP} block, incurring additional hardware and power overheads.

Regarding performance, the proposed \emph{Multispeculative PositMAC} achieves the highest operating frequency among all evaluated architectures, reaching \frequency{2.00} for both precisions. Compared to the \emph{Baseline}, our proposal reduces the per-cycle delay by $55.8\%$ for \ac{PositMAC32} and $64.3\%$ for \ac{PositMAC64}. When compared specifically to the fastest alternative supporting exact quire accumulation, \emph{Crespo MAC}, with frequencies of \frequency{1.11} and \frequency{0.49} for the 32- and 64-bit implementations, respectively, it achieves delay reductions of $44.4\%$ and $75.5\%$ for \ac{PositMAC32} and \ac{PositMAC64}, respectively.

In terms of area, despite incorporating a deeper pipeline and multispeculative logic, our proposal is strictly smaller than the \emph{Baseline}, which is the alternative implementing the quire with the lowest area consumption, achieving an area reduction of $19.5\%$ for the \ac{PositMAC32} unit and $17.6\%$ for the \ac{PositMAC64} unit. This reduction becomes more pronounced at higher precisions; for instance, the 64-bit \emph{Deep Positron} requires more than twice the area of our design. However, the non-quire implementation (\emph{Crespo FMA}) exhibits the lowest area consumption, despite implementing the \emph{Quire-to-Posit} encoder.

The architectural complexity and higher operating frequency of the proposed unit naturally lead to an increase in power consumption relative to the \emph{Baseline}, exceeding it by $35.5\%$ and $47.5\%$ for \ac{PositMAC32} and \ac{PositMAC64}, respectively. Nevertheless, this power overhead is heavily offset by the substantial reduction in execution time, making the \emph{Multispeculative PositMAC} highly energy-efficient and reducing per-cycle energy consumption by $40.1\%$ for \ac{PositMAC32} and $47.3\%$ for \ac{PositMAC64} relative to the \emph{Baseline}. While the architecture lacking quire integration (\emph{Crespo FMA}) inherently achieves the lowest per-cycle energy consumption for \ac{PositMAC32}, this trend reverses at higher precisions, with our proposal achieving a $0.9\%$ per-cycle energy saving over \emph{Crespo FMA} for \ac{PositMAC64}, thus delivering higher energy efficiency while simultaneously incorporating the quire accumulator.

When evaluating an execution of $L=40$ MAC operations, these per-cycle speedups directly dictate total performance and efficiency. The \emph{Multispeculative PositMAC} cuts the execution time by $30.9\%$ for \ac{PositMAC32} and $40.7\%$ for \ac{PositMAC64} compared to the \emph{Baseline}, while improving performance over the fastest alternatives, outperforming the 32-bit \emph{Crespo MAC} by $38.7\%$ and the 64-bit \emph{Crespo FMA} by $72.4\%$ in total execution time. In addition, this runtime reduction yields total execution energy savings of $30.86\%$ and $40.6\%$ for 32-bit and 64-bit formats relative to the \emph{Baseline}. Notably, for \ac{PositMAC64}, our unit also surpasses the non-quire implementation (\emph{Crespo FMA}) in execution delay, although it increases the energy consumption for the full execution sequence due to the additional cycles incurred during the speculation phase.


\section{Conclusions}
\label{sec:conclusions}

This work introduces the \emph{Multispeculative PositMAC}, a high-frequency variant of the baseline units optimized for energy efficiency. First, splitting the initial pipeline stage to balance critical delays reduces single-cycle energy consumption by up to $26.1\%$ in \ac{PositMAC64} with an area overhead of only up to $12.8\%$.

Second, the \emph{Multiplier} was identified as the primary performance bottleneck; evaluating multiple multiplier architectures revealed that the \emph{Booth-4/KS} implementation reduces critical-stage delay by up to $44.4\%$, satisfying the 0.5\,ns timing constraint and yielding modest area savings, albeit with increased total power consumption.

Finally, replacing the conventional monolithic accumulation adder with the proposed \emph{Multispeculative Adder} provides significant hardware savings, though it requires additional control circuitry. Even with this extra logic, the multispeculative designs retain substantial area and power advantages over their monolithic baselines. The lowest absolute area is achieved with $w=4$ using either \emph{Kogge--Stone} or \emph{Brent--Kung}, yielding reductions of up to $19.8\%$, depending on the adder architecture. Similarly, the lowest absolute power is achieved with \emph{Brent--Kung} at $w=256$, yielding power savings of up to $20.8\%$.

In comparison with the state-of-the-art, the proposed \emph{Multispeculative PositMAC} achieves the highest operating frequency and drastically cuts the critical cycle delay by up to $64.3\%$ relative to the \emph{Baseline}, and up to $79.0\%$ compared to quire-enabled alternatives. This performance is attained without increasing resource overhead, as the design remains strictly smaller in area and achieves lower per-cycle energy consumption than all quire-capable counterparts.

\ifCLASSOPTIONcompsoc
  \section*{Acknowledgments}
\else
  \section*{Acknowledgment}
\fi
This work was supported by grant PID2021-123041OB-I00 funded by MCIN/AEI/10.13039/501100011033 and by “ERDF A way of making Europe”, and by grant  PID2024-158203OB-I00 funded by MICIU/AEI/10.13039/501100011033 and by ERDF/EU. 

\ifCLASSOPTIONcaptionsoff
  \newpage
\fi

\bibliographystyle{IEEEtran}
\bibliography{bibliography}

@article{IEEEComputerSociety2008,
  author    = {{IEEE Computer Society}},
  doi       = {10.1109/IEEESTD.2008.4610935},
  isbn      = {9780738157528},
  journal   = {IEEE Std 754-2008 (Revision of IEEE 754-1985)},
  pages     = {1--70},
  publisher = {IEEE},
  title     = {{IEEE Standard for Floating-Point Arithmetic}},
  volume    = {2008},
  year      = {2008}
}

@ARTICLE{IEEE2019,
  author    = {{IEEE Computer Society}},
  journal={IEEE Std 754-2019 (Revision of IEEE 754-2008)}, 
  title={IEEE Standard for Floating-Point Arithmetic}, 
  year={2019},
  volume={},
  number={},
  pages={1-84},
  doi={10.1109/IEEESTD.2019.8766229}
}

@incollection{kulisch2008Computer,
  title = {Computer {{Arithmetic}} and {{Validity}}: {{Theory}}, {{Implementation}}, and {{Applications}}},
  shorttitle = {Computer {{Arithmetic}} and {{Validity}}},
  booktitle = {Computer {{Arithmetic}} and {{Validity}}},
  author = {Kulisch, Ulrich},
  year = {2008},
  month = aug,
  publisher = {De Gruyter},
  doi = {10.1515/9783110203196},
  isbn = {978-3-11-020319-6},
  langid = {english}
}

@article{dedinechin2011Designing,
  title = {Designing {{Custom Arithmetic Data Paths}} with {{FloPoCo}}},
  author = {De Dinechin, Florent and Pasca, Bogdan},
  year = {2011},
  month = jul,
  journal = {IEEE Design \& Test of Computers},
  volume = {28},
  number = {4},
  pages = {18--27},
  issn = {0740-7475},
  doi = {10.1109/MDT.2011.44},
  copyright = {https://ieeexplore.ieee.org/Xplorehelp/downloads/license-information/IEEE.html},
  langid = {english}
}

@article{Gustafson2017Beating,
  author    = {Gustafson, John L. and Yonemoto, Isaac},
  doi       = {10.14529/jsfi170206},
  issn      = {23138734},
  journal   = {Supercomputing Frontiers and Innovations},
  month     = {jun},
  number    = {2},
  pages     = {71--86},
  publisher = {South Ural State University, Publishing Center},
  title     = {{Beating Floating Point at its Own Game: Posit Arithmetic}},
  volume    = {4},
  year      = {2017}
}

@misc{positworkinggroup2022Standard,
  title = {Standard for {{Posit Arithmetic}} (2022)},
  shorttitle = {Standard for {{Posit Arithmetic}} (2022)},
  author = {{Posit Working Group}},
  year = {2022},
  month = feb,
  urldate = {2025-03-05},
  howpublished = {\url{https://posithub.org/docs/posit\_standard-2.pdf}}
}

@inproceedings{DeDinechin2019,
  address   = {New York, NY, USA},
  author    = {de Dinechin, Florent and
               Forget, Luc and
               Muller, Jean-Michel and
               Uguen, Yohann},
  booktitle = {2019 Conference for Next Generation Arithmetic (CoNGA)},
  doi       = {10.1145/3316279.3316285},
  isbn      = {9781450371391},
  month     = {mar},
  pages     = {1--10},
  publisher = {ACM},
  title     = {{Posits: the good, the bad and the ugly}},
  year      = {2019}
}

@inproceedings{Uguen2019,
  author    = {Uguen, Yohann and Forget, Luc and de Dinechin, Florent},
  booktitle = {2019 29th International Conference on Field Programmable Logic and Applications (FPL)},
  doi       = {10.1109/FPL.2019.00026},
  isbn      = {978-1-7281-4884-7},
  month     = {sep},
  pages     = {106--113},
  publisher = {IEEE},
  title     = {{Evaluating the Hardware Cost of the Posit Number System}},
  year      = {2019}
}

@inproceedings{Guntoro2020,
  author    = {Guntoro, Andre and {De La Parra}, Cecilia and Merchant, Farhad and {De Dinechin}, Florent and Gustafson, John L. and Langhammer, Martin and Leupers, Rainer and Nambiar, Sangeeth},
  booktitle = {2020 Design, Automation \& Test in Europe Conference \& Exhibition (DATE)},
  doi       = {10.23919/DATE48585.2020.9116196},
  isbn      = {978-3-9819263-4-7},
  month     = {mar},
  pages     = {1357--1365},
  publisher = {IEEE},
  title     = {{Next Generation Arithmetic for Edge Computing}},
  year      = {2020}
}

@incollection{zhang2024review,
  author    = {Zhang, Hao and Wei, Zhiqiang and Yin, Bo and Ko, Seok-Bum},
  title     = {{A Review of Posit Arithmetic for Energy-Efficient Computation: Methodologies, Applications, and Challenges}},
  booktitle = {Design and Applications of Emerging Computer Systems},
  year      = {2024},
  publisher = {Springer Nature Switzerland},
  pages     = {649--670},
  isbn      = {978-3-031-42478-6},
  doi       = {10.1007/978-3-031-42478-6_24},
  url       = {https://doi.org/10.1007/978-3-031-42478-6\_24}
}

@article{mallasen2025navigating,
  author = {Mallas\'{e}n, David and Murillo, Raul and Botella, Guillermo and Del Barrio, Alberto Antonio},
  title = {Navigating Posit Arithmetic: A Comprehensive Survey of Principles, Hardware, and Applications},
  year = {2025},
  issue_date = {April 2026},
  publisher = {Association for Computing Machinery},
  address = {New York, NY, USA},
  volume = {58},
  number = {5},
  issn = {0360-0300},
  url = {https://doi.org/10.1145/3772284},
  doi = {10.1145/3772284},
  journal = {ACM Comput. Surv.},
  month = nov,
  articleno = {131},
  numpages = {36}
}

@inproceedings{Zhang2019,
  author    = {Zhang, Hao and He, Jiongrui and Ko, Seok-Bum},
  booktitle = {2019 IEEE International Symposium on Circuits and Systems (ISCAS)},
  doi       = {10.1109/ISCAS.2019.8702349},
  isbn      = {978-1-7281-0397-6},
  month     = {may},
  pages     = {1--5},
  publisher = {IEEE},
  title     = {{Efficient Posit Multiply-Accumulate Unit Generator for Deep Learning Applications}},
  volume    = {2019-May},
  year      = {2019}
}

@inproceedings{Murillo2020Customized,
  author    = {Murillo, Raul and {Del Barrio}, Alberto A. and Botella, Guillermo},
  booktitle = {2020 IEEE International Symposium on Circuits and Systems (ISCAS)},
  doi       = {10.1109/iscas45731.2020.9180771},
  isbn      = {978-1-7281-3320-1},
  month     = {oct},
  pages     = {1--5},
  publisher = {IEEE},
  title     = {{Customized Posit Adders and Multipliers using the FloPoCo Core Generator}},
  year      = {2020}
}

@article{Neves2020,
  author  = {Neves, Nuno and Tomas, Pedro and Roma, Nuno},
  doi     = {10.1109/SiPS50750.2020.9195256},
  isbn    = {978-1-7281-8099-1},
  journal = {2020 IEEE Workshop on Signal Processing Systems (SiPS)},
  pages   = {1--6},
  title   = {{Dynamic Fused Multiply-Accumulate Posit Unit with Variable Exponent Size for Low-Precision DSP Applications}},
  year    = {2020}
}

@inproceedings{murillo2021EnergyEfficient,
  title = {Energy-{{Efficient MAC Units}} for {{Fused Posit Arithmetic}}},
  booktitle = {2021 {{IEEE}} 39th {{International Conference}} on {{Computer Design}} ({{ICCD}})},
  author = {Murillo, Ra{\'u}l and Mallas{\'e}n, David and Del Barrio, Alberto A. and Botella, Guillermo},
  year = {2021},
  month = oct,
  pages = {138--145},
  issn = {2576-6996},
  doi = {10.1109/ICCD53106.2021.00032}
}

@incollection{murillo2022Comparing,
  title = {Comparing {{Different Decodings}} for {{Posit Arithmetic}}},
  booktitle = {Next {{Generation Arithmetic}}},
  author = {Murillo, Raul and Mallas{\'e}n, David and Del Barrio, Alberto A. and Botella, Guillermo},
  editor = {Gustafson, John L. and Dimitrov, Vassil},
  year = {2022},
  volume = {13253},
  pages = {84--99},
  publisher = {Springer International Publishing},
  address = {Cham},
  doi = {10.1007/978-3-031-09779-9_6},
  isbn = {978-3-031-09778-2 978-3-031-09779-9},
  langid = {english}
}

@article{crespo2022Unified,
  title = {Unified {{Posit}}/{{IEEE-754 Vector MAC Unit}} for {{Transprecision Computing}}},
  author = {Crespo, Luis and Tomas, Pedro and Roma, Nuno and Neves, Nuno},
  year = {2022},
  month = may,
  journal = {IEEE Transactions on Circuits and Systems II: Express Briefs},
  volume = {69},
  number = {5},
  pages = {2478--2482},
  issn = {1549-7747, 1558-3791},
  doi = {10.1109/TCSII.2022.3160191},
  copyright = {https://ieeexplore.ieee.org/Xplorehelp/downloads/license-information/IEEE.html},
  langid = {english}
}

@inproceedings{ledoux2022Generator,
  title = {A {{Generator}} of {{Numerically-Tailored}} and {{High-Throughput Accelerators}} for {{Batched GEMMs}}},
  booktitle = {2022 {{IEEE}} 30th {{Annual International Symposium}} on {{Field-Programmable Custom Computing Machines}} ({{FCCM}})},
  author = {Ledoux, Louis and Casas, Marc},
  year = {2022},
  month = may,
  pages = {1--10},
  publisher = {IEEE},
  address = {New York City, NY, USA},
  doi = {10.1109/FCCM53951.2022.9786164},
  copyright = {https://doi.org/10.15223/policy-029},
  isbn = {978-1-66548-332-2},
  langid = {english}
}

@inproceedings{nakasato2024Evaluation,
  title = {Evaluation of {{POSIT Arithmetic}} with {{Accelerators}}},
  booktitle = {Proceedings of the {{International Conference}} on {{High Performance Computing}} in {{Asia-Pacific Region}}},
  author = {Nakasato, Naohito and Murakami, Yuki and Kono, Fumiya and Nakata, Maho},
  year = {2024},
  month = jan,
  pages = {62--72},
  publisher = {ACM},
  address = {Nagoya Japan},
  doi = {10.1145/3635035.3635046},
  isbn = {9798400708893},
  langid = {english}
}

@article{neves2021Reconfigurable,
  title = {A {{Reconfigurable Posit Tensor Unit}} with {{Variable-Precision Arithmetic}} and {{Automatic Data Streaming}}},
  author = {Neves, Nuno and Tom{\'a}s, Pedro and Roma, Nuno},
  year = {2021},
  month = dec,
  journal = {Journal of Signal Processing Systems},
  volume = {93},
  number = {12},
  pages = {1365--1385},
  issn = {1939-8018, 1939-8115},
  doi = {10.1007/s11265-021-01687-7},
  langid = {english}
}

@inproceedings{li2023PDPU,
  title = {{{PDPU}}: {{An Open-Source Posit Dot-Product Unit}} for {{Deep Learning Applications}}},
  shorttitle = {{{PDPU}}},
  booktitle = {2023 {{IEEE International Symposium}} on {{Circuits}} and {{Systems}} ({{ISCAS}})},
  author = {Li, Qiong and Fang, Chao and Wang, Zhongfeng},
  year = {2023},
  month = may,
  pages = {1--5},
  issn = {2158-1525},
  doi = {10.1109/ISCAS46773.2023.10182007}
}

@article{murillo2023Generating,
  title = {Generating {{Posit-Based Accelerators With High-Level Synthesis}}},
  author = {Murillo, Raul and Barrio, Alberto A. Del and Botella, Guillermo and Pilato, Christian},
  year = {2023},
  month = oct,
  journal = {IEEE Transactions on Circuits and Systems I: Regular Papers},
  volume = {70},
  number = {10},
  pages = {4040--4052},
  issn = {1549-8328, 1558-0806},
  doi = {10.1109/TCSI.2023.3299009},
  copyright = {https://creativecommons.org/licenses/by-nc-nd/4.0/},
  langid = {english}
}

@misc{wu2025pvu,
  title={PVU: Design and Implementation of a Posit Vector Arithmetic Unit (PVU) for Enhanced Floating-Point Computing in Edge and AI Applications}, 
  author={Xinyu Wu and Yaobin Wang and Tianyi Zhao and Jiawei Qin and Zhu Liang and Jie Fu},
  year={2025},
  eprint={2503.01313},
  archivePrefix={arXiv},
  primaryClass={cs.DC},
  url={https://arxiv.org/abs/2503.01313}, 
}

@INPROCEEDINGS{crespo2023trading,
  author={Crespo, Luís and Tomás, Pedro and Roma, Nuno and Neves, Nuno},
  booktitle={2023 IEEE 35th International Symposium on Computer Architecture and High Performance Computing (SBAC-PAD)}, 
  title={Trading Performance, Power, and Area on Low-Precision Posit MAC Units for CNN Training}, 
  year={2023},
  volume={},
  number={},
  pages={46-56},
  doi={10.1109/SBAC-PAD59825.2023.00014}
}

@ARTICLE{condia2025investigating,
  author={Condia, Josie E. Rodriguez and Guerrero-Balaguera, Juan-David and Sierra, Robert Limas and Reorda, Matteo Sonza},
  journal={IEEE Transactions on Emerging Topics in Computing}, 
  title={Investigating and Mitigating Critical Faults in Floating-Point and Posit Arithmetic Hardware}, 
  year={2025},
  volume={},
  number={},
  pages={1-12},
  doi={10.1109/TETC.2025.3615827}
}

@INPROCEEDINGS{dube2025compact,
  author={Dube, Ayushi and Singh, Gian and Vrudhula, Sarma},
  booktitle={2025 IEEE/ACM International Symposium on Low Power Electronics and Design (ISLPED)}, 
  title={A Compact, Low Power Transprecision ALU for Smart Edge Devices}, 
  year={2025},
  volume={},
  number={},
  pages={1-8},
  doi={10.1109/ISLPED65674.2025.11261699}
}

@ARTICLE{kumar2026spade,
       author = {{Kumar}, Sonu and {Vinnakota}, Lavanya and {Lokhande}, Mukul and {Vishvakarma}, Santosh Kumar and {Teman}, Adam},
        title = "{SPADE: A SIMD Posit-enabled compute engine for Accelerating DNN Efficiency}",
      journal = {arXiv e-prints},
         year = 2026,
        month = jan,
          eid = {arXiv:2601.17279},
        pages = {arXiv:2601.17279},
          doi = {10.48550/arXiv.2601.17279},
archivePrefix = {arXiv},
       eprint = {2601.17279},
 primaryClass = {cs.AR},
       adsurl = {https://ui.adsabs.harvard.edu/abs/2026arXiv260117279K}
}

@inproceedings{Carmichael2019a,
  author    = {Carmichael, Zachariah and Langroudi, Hamed F. and Khazanov, Char and Lillie, Jeffrey and Gustafson, John L. and Kudithipudi, Dhireesha},
  booktitle = {2019 Design, Automation \& Test in Europe Conference \& Exhibition (DATE)},
  doi       = {10.23919/DATE.2019.8715262},
  isbn      = {978-3-9819263-2-3},
  month     = {mar},
  pages     = {1421--1426},
  publisher = {IEEE},
  title     = {{Deep Positron: A Deep Neural Network Using the Posit Number System}},
  year      = {2019}
}

@article{lu2021Evaluations,
  title = {Evaluations on {{Deep Neural Networks Training Using Posit Number System}}},
  author = {Lu, Jinming and Fang, Chao and Xu, Mingyang and Lin, Jun and Wang, Zhongfeng},
  year = {2021},
  month = feb,
  journal = {IEEE Transactions on Computers},
  volume = {70},
  number = {2},
  pages = {174--187},
  issn = {1557-9956},
  doi = {10.1109/TC.2020.2985971}
}

@article{Murillo2020Deep,
  author   = {Murillo, Raul and {Del Barrio}, Alberto A. and Botella, Guillermo},
  doi      = {10.1016/j.dsp.2020.102762},
  issn     = {10512004},
  journal  = {Digital Signal Processing: A Review Journal},
  month    = {jul},
  pages    = {102762},
  title    = {{Deep PeNSieve: A deep learning framework based on the posit number system}},
  volume   = {102},
  year     = {2020}
}

@inproceedings{ramachandran2024AlgorithmHardware,
  author = {Ramachandran, Akshat and Wan, Zishen and Jeong, Geonhwa and Gustafson, John L. and Krishna, Tushar},
  title = {Algorithm-Hardware Co-Design of Distribution-Aware Logarithmic-Posit Encodings for Efficient DNN Inference},
  year = {2024},
  isbn = {9798400706011},
  publisher = {Association for Computing Machinery},
  address = {New York, NY, USA},
  doi = {10.1145/3649329.3656544},
  booktitle = {Proceedings of the 61st ACM/IEEE Design Automation Conference},
  articleno = {326},
  numpages = {6},
  location = {San Francisco, CA, USA},
  series = {DAC '24}
}

@INPROCEEDINGS{crafton2025finding,
  author={Crafton, Brian and Peng, Xiaochen and Sun, Xiaoyu and Lele, Ashwin and Zhang, Bo and Khwa, Win-San and Akarvardar, Kerem},
  booktitle={2025 62nd ACM/IEEE Design Automation Conference (DAC)}, 
  title={Finding the Pareto Frontier of Low-Precision Data Formats and MAC Architecture for LLM Inference}, 
  year={2025},
  volume={},
  number={},
  pages={1-7},
  doi={10.1109/DAC63849.2025.11132989}
}

@ARTICLE{hao2025positCIM,
  author={Wu, Hao and Chen, Yong and Li, Ming and Zhang, Bingxin and Zhang, Rui and Yuan, Yiyang and Yang, Yiming and Yue, Jinshan and Wang, Xinghua and Li, Xiaoran and Zhang, Feng},
  journal={IEEE Journal of Solid-State Circuits}, 
  title={A 28-nm 88.3-TFLOPS/W POSIT-Approximate-Calculation-Based Digital Computing-in-Memory Macro Incorporating Final-Cycle Fusion and Joint Skipping}, 
  year={2025},
  volume={},
  number={},
  pages={1-15},
  doi={10.1109/JSSC.2025.3590632}
}

@ARTICLE{prabhu2025minotaur,
  author={Prabhu, Kartik and Radway, Robert M. and Yu, Jeffrey and Bartolone, Kai and Giordano, Massimo and Peddinghaus, Fabian and Urman, Yonatan and Khwa, Win-San and Chih, Yu-Der and Chang, Meng-Fan and Mitra, Subhasish and Raina, Priyanka},
  journal={IEEE Journal of Solid-State Circuits}, 
  title={MINOTAUR: A Posit-Based 0.42–0.50-TOPS/W Edge Transformer Inference and Training Accelerator}, 
  year={2025},
  volume={60},
  number={4},
  pages={1311-1323},
  doi={10.1109/JSSC.2025.3545731}
}

@misc{mallasen2025phee,
      title={Increasing the Energy-Efficiency of Wearables Using Low-Precision Posit Arithmetic with PHEE}, 
      author={David Mallasén and Pasquale Davide Schiavone and Alberto A. Del Barrio and Manuel Prieto-Matias and David Atienza},
      year={2025},
      eprint={2501.18253},
      archivePrefix={arXiv},
}

@inproceedings{arunkumar2020PERC,
  title = {{{PERC}}: {{Posit Enhanced Rocket Chip}}},
  booktitle = {4th {{Workshop}} on {{Computer Architecture Research}} with {{RISC-V}} ({{CARRV}}'20)},
  author = {Arunkumar, M. V. and Bhairathi, Sai Ganesh and Hayatnagarkar, Harshal G.},
  year = {2020},
  pages = {8},
  langid = {english}
}

@article{Tiwari2021,
  author   = {Tiwari, Sugandha and Gala, Neel and Rebeiro, Chester and Kamakoti, V.},
  doi      = {10.1145/3446210},
  issn     = {1544-3566},
  journal  = {ACM Transactions on Architecture and Code Optimization},
  month    = {apr},
  number   = {3},
  pages    = {1--26},
  title    = {{PERI: A Configurable Posit Enabled RISC-V Core}},
  volume   = {18},
  year     = {2021}
}

@article{cococcioni2022Lightweight,
  title = {A {{Lightweight Posit Processing Unit}} for {{RISC-V Processors}} in {{Deep Neural Network Applications}}},
  author = {Cococcioni, Marco and Rossi, Federico and Ruffaldi, Emanuele and Saponara, Sergio},
  year = {2022},
  month = oct,
  journal = {IEEE Transactions on Emerging Topics in Computing},
  volume = {10},
  number = {4},
  pages = {1898--1908},
  issn = {2168-6750, 2376-4562},
  doi = {10.1109/TETC.2021.3120538},
  copyright = {https://ieeexplore.ieee.org/Xplorehelp/downloads/license-information/IEEE.html},
  langid = {english}
}

@article{sharma2023CLARINET,
  title = {{{CLARINET}}: {{A}} Quire-Enabled {{RISC-V-based}} Framework for Posit Arithmetic Empiricism},
  shorttitle = {{{CLARINET}}},
  author = {Sharma, Niraj N. and Jain, Riya and Pokkuluri, Mohana Madhumita and Patkar, Sachin B. and Leupers, Rainer and Nikhil, Rishiyur S. and Merchant, Farhad},
  year = {2023},
  month = feb,
  journal = {Journal of Systems Architecture},
  volume = {135},
  pages = {102801},
  issn = {13837621},
  doi = {10.1016/j.sysarc.2022.102801},
  langid = {english}
}

@article{mallasen2022PERCIVAL,
  title = {{{PERCIVAL}}: {{Open-Source Posit RISC-V Core With Quire Capability}}},
  shorttitle = {{{PERCIVAL}}},
  author = {Mallas{\'e}n, David and Murillo, Raul and Barrio, Alberto A. Del and Botella, Guillermo and Pi{\~n}uel, Luis and {Prieto-Matias}, Manuel},
  year = {2022},
  month = jul,
  journal = {IEEE Transactions on Emerging Topics in Computing},
  volume = {10},
  number = {3},
  pages = {1241--1252},
  issn = {2168-6750, 2376-4562},
  doi = {10.1109/TETC.2022.3187199},
  langid = {english}
}

@article{mallasen2024BigPERCIVAL,
  title = {Big-{{PERCIVAL}}: {{Exploring}} the {{Native Use}} of 64-{{Bit Posit Arithmetic}} in {{Scientific Computing}}},
  shorttitle = {Big-{{PERCIVAL}}},
  author = {Mallas{\'e}n, David and Del Barrio, Alberto A. and {Prieto-Matias}, Manuel},
  year = {2024},
  month = jun,
  journal = {IEEE Transactions on Computers},
  volume = {73},
  number = {6},
  pages = {1472--1485},
  issn = {0018-9340, 1557-9956, 2326-3814},
  doi = {10.1109/TC.2024.3377890},
  copyright = {https://creativecommons.org/licenses/by-nc-nd/4.0/},
  langid = {english}
}

@ARTICLE{delbarrio2012multispeculative,
  author={Del Barrio, Alberto A. and Hermida, Román and Memik, Seda Ogrenci and Mendias, José M. and Molina, María C.},
  journal={IEEE Transactions on Computer-Aided Design of Integrated Circuits and Systems}, 
  title={Multispeculative Addition Applied to Datapath Synthesis}, 
  year={2012},
  volume={31},
  number={12},
  pages={1817-1830},
  doi={10.1109/TCAD.2012.2208966}
}

@ARTICLE{delbarrio2019combined,
  author={Del Barrio, Alberto A. and Hermida, Román and Ogrenci-Memik, Seda},
  journal={IEEE Transactions on Circuits and Systems I: Regular Papers}, 
  title={A Combined Arithmetic-High-Level Synthesis Solution to Deploy Partial Carry-Save Radix-8 Booth Multipliers in Datapaths}, 
  year={2019},
  volume={66},
  number={2},
  pages={742-755},
  doi={10.1109/TCSI.2018.2866172}
}

@ARTICLE{delbarrio2016partial,
  author={Barrio, Alberto A. Del and Hermida, Román and Memik, Seda Ogrenci},
  journal={IEEE Transactions on Computers}, 
  title={A Partial Carry-Save On-the-Fly Correction Multispeculative Multiplier}, 
  year={2016},
  volume={65},
  number={11},
  pages={3251-3264},
  doi={10.1109/TC.2016.2529626}
}

@misc{machetti2024xheep,
  author       = {Simone Machetti and
                  Pasquale D. Schiavone and
                  Thomas C. M{\"{u}}ller and
                  Miguel Pe{\'{o}}n and
                  David Atienza},
  title        = {{X-HEEP:} An Open-Source, Configurable and Extendible {RISC-V} Microcontroller
                  for the Exploration of Ultra-Low-Power Edge Accelerators},
  year         = {2024},
  eprinttype    = {arXiv},
  eprint       = {2401.05548},
}

@article{kogge1973parallel,
  author = {Kogge, Peter M. and Stone, Harold S.},
  title = {A Parallel Algorithm for the Efficient Solution of a General Class of Recurrence Equations},
  year = {1973},
  issue_date = {August 1973},
  publisher = {IEEE Computer Society},
  address = {USA},
  volume = {22},
  number = {8},
  issn = {0018-9340},
  url = {https://doi.org/10.1109/TC.1973.5009159},
  doi = {10.1109/TC.1973.5009159},
  journal = {IEEE Trans. Comput.},
  month = aug,
  pages = {786–793},
  numpages = {8}
}

@article{booth1951signed,
    author = {Booth, Andrew D.},
    title = {A Signed Binary Multiplication Technique},
    journal = {The Quarterly Journal of Mechanics and Applied Mathematics},
    volume = {4},
    number = {2},
    pages = {236-240},
    year = {1951},
    month = {01},
    issn = {0033-5614},
    doi = {10.1093/qjmam/4.2.236},
    url = {https://doi.org/10.1093/qjmam/4.2.236},
    eprint = {https://academic.oup.com/qjmam/article-pdf/4/2/236/5301697/4-2-236.pdf},
}

@book{Ercegovac2004,
  author    = {Miloš D. Ercegovac and Tomás Lang},
  pages     = {709},
  publisher = {Elsevier},
  title     = {{Digital Arithmetic}},
  year      = {2004},
  doi       = {10.1016/B978-1-55860-798-9.X5000-3}
}




\end{document}